\documentclass[12pt]{article}
\usepackage{}
\usepackage{geometry}             
\usepackage{graphicx}
\usepackage{amssymb}
\usepackage{amsmath}
\usepackage{epstopdf}
\usepackage{comment}
\usepackage{cite}
\usepackage{caption}
\usepackage{subfigure}

\usepackage[hyperindex=true,
          pdfstartview=FitH,
          bookmarksnumbered=true,
          bookmarksopen=true,
          citecolor=blue,
          linkcolor=blue,
          colorlinks=true,
          pdfborder=001,
          unicode]{hyperref}

\allowdisplaybreaks

\begin{document}

\title{$(2+1)$-dimensional charged black holes with scalar hair in Einstein-Power-Maxwell Theory}
\author{Wei Xu$^{1}$ \thanks{Correspondence author} \thanks{{\em
        email}: \href{mailto:xuweifuture@gmail.com}
        {xuweifuture@gmail.com}}
         and De-Cheng Zou$^{2}$ \thanks{{\em
        email}: \href{mailto:zoudecheng@sjtu.edu.cn}
        {zoudecheng@sjtu.edu.cn}}\\
$^{1}$School of Physics, Huazhong University of Science and Technology, \\ Wuhan 430074, China\\
$^{2}$Department of Physics and Astronomy, Shanghai Jiao Tong University, \\Shanghai 200240, China}
\date{}
\maketitle

\begin{abstract}
We obtain an exact static solution to Einstein-Power-Maxwell (EPM) theory in
$(2+1)$ dimensional AdS spacetime, in which the scalar field couples to gravity in a
non-minimal way. After considering the scalar potential, a stable system leads to a constraint on the power parameter $k$ of Maxwell field. The solution contains a curvature singularity at the origin and is non-conformally flat. The
horizon structures are identified, which indicates the physically acceptable lower bound of mass in according to the existence of black hole solutions. Especially for the cases with $k>1$, the lower bound is negative, thus there exist scalar black holes with negative mass.
The null geodesics in this spacetime are also discussed in detail. They are divided into five models,
which are made up of the cases with the following geodesic motions: no-allowed motion, the circular motion,
the elliptic motion and the unbounded spiral motion.

\vspace{3mm}

\vspace{3mm}


\end{abstract}

\section{Introduction}

The no-hair theorems assert that the asymptotically flat spacetime can not admit any hairy
black hole solutions \cite{Bocharova,Bekenstein:1974sf,Bekenstein:1975ts},
since the scalar field was divergent on the horizon and stability
analysis showed that they were unstable \cite{Bronnikov:1978mx}.
Nevertheless, when a negative cosmological constant is considered, the no-hair theorems
can usually be circumvented, and then there exists a broad literature of black hole solutions
with a (non)minimal scalar field in the four or higher dimensional Einstein's gravity,
including static \cite{Martinez:2004nb,Martinez:2006an,Martinez:2005di,Kolyvaris:2009pc,
Gonzalez:2013aca,Nadalini:2007qi,Feng:2013tza,Acena:2012mr,Acena:2013jya,Anabalon:2013sra}, and rotating \cite{Herdeiro:2014goa,Herdeiro:2014jaa,Anabalon:2012dw}
extensions with a complex, massive scalar field and in
higher order derivative gravity \cite{Gaete:2013ixa,Correa:2013bza,Gaete:2013oda,Giribet:2014bva}.

Inspiring by the pioneering work of Banados, Teitelboim and Zanelli (BTZ) \cite{Banados:1992wn},
$(2+1)$-dimensional spacetimes admitting black hole solutions
have attracted much attention. Besides sourced by a mass and a negative cosmological constant,
pure BTZ black holes can be also to added new sources such as electric/magnetic fields from
Maxwell's theory \cite{Cataldo:1999wr,Mazharimousavi:2011nd,Gurtug:2010dr},
Maxwell-dilaton \cite{Chan:1994qa}, rotation \cite{Martinez:1999qi},
perfect fluid \cite{Astorino:2011mw,Xu:2011vp} and others \cite{Garcia:2002rn,Wu:2013wca}.
It is interesting to note that the nonlinear electrodynamics (NED) is useful to overcome to
eliminate the electromagnetic singularities due to point charges that occur
in the linear Maxwell theory. Besides, the nonlinear electrodynamics has been considered in general relativity, which are a good laboratories in order to construct black hole solutions \cite{Hassaine:2007py,Hassaine:2008pw,Maeda:2008ha}. Black hole solutions with nonlinear electrodynamics sources have interesting asymptotic behaviors and exhibit interesting thermodynamics properties \cite{Gonzalez:2009nn,DiazAlonso:2012mb,Bazrafshan:2012rn,Arciniega:2014iya}. For example, they satisfy the zeroth and first laws of black-hole mechanics \cite{Rasheed:1997ns}. After considering the cosmological constant as a dynamical pressure, the Smarr relation work as well and there are rich phase structure which have the first order phase transitions and the reentrant phase transitions \cite{Hendi:2012um,Mo:2014qsa}. 

On the other hand, with a negative cosmological constant in the action,
the $(2+1)$-dimensional black holes with the minimal \cite{Banados:2005hm,Henneaux:2002wm,Schmidt:2012pp}
or non-minimal \cite{Martinez:1996gn,Hortacsu:2003we} scalar fields have been constructed
in the Einstein's gravity. Furthermore, the charged \cite{Xu:2013nia},
rotating \cite{Zhao:2013isa,Degura:1998hw,Aparicio:2012yq,Zou:2014gla}, charged rotating \cite{Sadeghi:2013gmf} and
Einstein-Born-Infeld \cite{Mazharimousavi:2014vza} black holes with non-minimally
scalar hair and rotating black hole dressed with minimal scalar field hair \cite{Xu:2014uha}
in the $(2+1)$ dimensional Einstein's gravity, and black hole dressed by a (non)minimally
scalar field in New Massive Gravity \cite{Correa:2014ika} have been discussed.
Beyond the linear Maxwell electromagnetism in theory with scalar fields, in this paper, we study
the charged black hole solution with non-minimally
coupled scalar field in the $(2+1)$-dimensional Einstein-Power-Maxwell (EPM) theory.
Actually, asymptotically AdS black holes with nonlinear electrodynamics sources endowed 
with a extra scalar field have been related to superconductors by means of the gravity/gauge
duality \cite{Jing:2011vz,Jing:2012dj,Banerjee:2012vk,Roychowdhury:2012vj,Dey:2013qoa}. Especially, the larger power parameter $k$ of the Power Maxwell field makes it harder for the scalar hair to be condensated \cite{Jing:2011vz}. This makes it more interesting to study the black hole solutions in this paper.
In addition, we will present the null geodesics in detail, in order to have a further understanding of the properties of this solution.

This paper is organized as follows. In Sec.~\ref{2s}, we present the charged black
hole solution in the EPM gravity with non-minimally coupled scalar field, and then discuss
the properties of the scalar potential.
Moreover, the basic geometric properties and horizon structure of the metric are also outlined.
In Sec.~\ref{3s} the geodesics motions are given for the photon.
The Sec.~\ref{4s} is devoted to the closing remarks.

\section{Charged black hole in the EPM theory with non-minimally coupled scalar field}
\label{2s}

\subsection{Charged hairy black hole solution}

The $(2+1)$-dimensional action in the Einstein-Power-Maxwell (EPM) theory with non-minimally
coupled scalar field is written as
\begin{align}
I=\int \mathrm{d}^{3}x\sqrt{-g}\left[\frac{1}{2\pi}\bigg(
R- \nabla_{\mu} \phi\nabla^{\mu} \phi -\frac{1}{8} R\phi^2-2V(\phi )\bigg)-L(\mathcal {F})\right] ,
\end{align}
in which $R$ is the Ricci scalar, $\phi$ is the scalar field,
$V(\phi)$ is the self-coupling potential of scalar field, and $L(\mathcal {F})=\mathcal {F}^k$
is the power Maxwell Lagrangian with the Maxwell invariant $\mathcal {F}=F_{\mu\nu}F^{\mu\nu}$.
Here power parameter $k$ is a rational number whose range is $k>\frac{1}{2}$,
restricted by the weak energy conditions (WEC) and strong energy
conditions (SEC) \cite{Gurtug:2010dr,Chan:2008gv}.
For $k=1$, the theory reduces to Einstein-Maxwell theory with non-minimally coupled
scalar field \cite{Xu:2013nia}. We does not consider this case in this paper.
Besides, the theory with $k=\frac{3}{4}$, in which the Maxwell source is conformally invariant, is presented in \cite{Cardenas} a few days ago.

Considering the variation of the action, we can obtain the field equations
\begin{align}
  &G^{\mu}{}_{\nu}-\pi\,T_{[A]}^{\mu}{}_{\nu}-T_{[\phi]}^{\mu}{}_{\nu}
  +V(\phi)\delta^{\mu}{}_{\nu}=0,\label{metric-eq}\\
  &\frac{1}{\sqrt{-g}}\partial_{\nu}(\sqrt{-g}F^{\mu\nu}\mathcal {F}^{k-1})=0,\label{maxwell-eq}\\
  &\nabla_{\mu}\nabla^{\mu}\phi-\frac{1}{8}R\phi-\partial_{\phi}V(\phi)=0,\label{scalar-eq}
\end{align}
where the energy-momentum tensor of the power Maxwell field and scalar field are given by
\begin{align}
  &T_{[A]}^{\mu}{}_{\nu}=\frac{\mathcal {F}^{k}}{2}\left(\frac{4k(F_{\nu\sigma}F^{\mu\sigma})}{\mathcal {F}}-\delta^{\mu}{}_{\nu}\right),\\
  &T_{[\phi]}^{\mu}{}_{\nu}=\partial^{\mu}\phi\partial_{\nu}\phi
  -\frac{1}{2}\delta^{\mu}{}_{\nu}\nabla^{\rho}\phi\nabla_{\rho}\phi
  +\frac{1}{8}\big(\delta^{\mu}{}_{\nu}\square-\nabla^{\mu}\nabla_{\nu}+G^{\mu}{}_{\nu}
  \big)\phi^2.
\end{align}

The metric ansatz is chosen as
\begin{align}
  \mathrm{d}s^2=-f(r)\mathrm{d}t^2+\frac{1}{f(r)}\mathrm{d}r^2+r^2\mathrm{d}\theta^2
\end{align}
with the coordinate range $-\infty<t<+\infty, r>0$ and $0<\theta<2\pi$.
In this setting, for $k\neq 1$, the vector potential of the power Maxwell field is given by the Maxwell equation
\begin{align}
A_{\mu}\mathrm{d}x^{\mu}=\frac{(2k-1)}{2(k-1)}Qr^{\frac{2(k-1)}{(2k-1)}}\mathrm{d}t,
\end{align}
where the parameter $Q$ is related to the electric charge $Q_{e}$ of black hole in the EPM theory \cite{Gurtug:2010dr}
\begin{align*}
Q_{e}=\frac{\sqrt{2}}{4}\pi k(2k-1)^{\frac{2k-1}{2k}}Q^{2k-1}.
\end{align*}
In the following paper, we will still use the parameter $Q$ for simplicity other than $Q_{e}$.

Then, we can obtain the following solution from the other field equations with the simplified form of scalar field shown in \cite{Xu:2013nia,Zhao:2013isa}
\begin{align}
  \phi(r)=\pm\sqrt{\frac{8B}{r+B}},\label{scalar-sol}
\end{align}
the metric function
\begin{align}
  f(r)=\frac{r^2}{\ell^2}+\frac{\alpha B^2(2B+3r)}{48r}+\frac{(-2)^{k}(2k-1)^2\pi \left(qB^{\frac{1}{(2k-1)}}\right)^{2k}}{2(k-1)}\frac{\left(r+\frac{2kB}{(4k-1)}\right)}{r^{\frac{1}{(2k-1)}}}\label{metric-sol}
\end{align}
and self-coupling potential of scalar field
\begin{align}
  V(\phi)&=-\frac{1}{\ell^2}+\frac{1}{512}\left(\frac{1}{\ell^2}+\frac{\alpha}{48}\right)\phi^6
  +\frac{(-2)^{k}(2k-1)\pi\,q^{2k}}{1024(k-1)(4k-1)}\left(\frac{\phi^2}{8-\phi^2}\right)^{\frac{2k}{(2k-1)}}\nonumber\\
  &\times \bigg(64k\phi^2-16k(2k-1)\phi^4+(2k-1)^2\phi^6\bigg),\label{potential-sol}
\end{align}
where the constant $q=\frac{Q}{B^{\frac{1}{2k-1}}}$ makes
the action of the theory invariant.  $\Lambda=V(0)=-\frac{1}{\ell^2}$ is the constant term emerging naturally in the potential
which plays the role of cosmological constant. The negative cosmological constant is necessary
for obtaining black hole solutions because of the No-Go theorem in three dimensions \cite{Ida:2000jh}.
The parameter $\alpha$ is related to the mass of black hole. Notice that the Maxwell field and self-coupling potential are singular when $k=1$. Actually, For $k=1$, the Maxwell vector potential reaches to the usual $\ln(r)$ form in three dimensions.

For the scalar field, the Eq.(\ref{scalar-sol}) shows that the potential $V(\phi)$ always
keeps regular when $\phi^2<8$. In order to obtain a stable system, the potential
should be bounded from below. Firstly, when $\phi^2\rightarrow8^{-}$, the scalar potential
\begin{align*}
  V(\phi)=\frac{\alpha}{48}-\frac{(2k-1)(-2)^k\pi{q}^{2k}}{2(4k-1)}\lim_{\phi^2\rightarrow8^{
  -}}\frac{1}{\left(1-\frac{\phi^2}{8}\right)^{\frac{2k}{(2k-1)}}}\simeq-(-1)^k\times(+\infty)
\end{align*}
must be positive, hence
\begin{align}
  (-1)^k=-1
  \label{k-value}
\end{align}
must holds. This leads to $k=\frac{2a+1}{2b+1}$ or $k=2c+1$,
where $a,b,c$ are all positive integer. Note for the case with conformally invariant Maxwell source, i.e. $k=\frac{3}{4}$, presented in \cite{Cardenas} a few days ago, is clear out of this constraint Eq.(\ref{k-value}), for the reason that the authors begin with the general form of the scalar field other than the simplified one Eq.(\ref{scalar-sol}).
Then, the horizon function Eq.~(\ref{metric-sol}) becomes
\begin{align}
  f(r)=\frac{r^2}{\ell^2}+\frac{\alpha B^2(2B+3r)}{48r}
  -\frac{2^{k}(2k-1)^2\pi q^{2k}\left(B^{\frac{2k}{(2k-1)}}\right)}{2(k-1)}\frac{\left(r
  +\frac{2kB}{(4k-1)}\right)}{r^{\frac{1}{(2k-1)}}}.\label{metric-sol-real}
\end{align}
Meanwhile, one can rewrite the potential as
\begin{align}
  V(\phi)&=-\frac{1}{\ell^2}+\frac{1}{512}\left(\frac{1}{\ell^2}+\frac{\alpha}{48}\right)\phi^6
  -\frac{2^{k}(2k-1)\pi\,q^{2k}}{1024(k-1)(4k-1)}\left(\frac{\phi^2}{8-\phi^2}\right)^{\frac{2k}{(2k-1)}}\nonumber\\
  &\times \bigg(64k\phi^2
  -16k(2k-1)\phi^4+(2k-1)^2\phi^6\bigg).\label{potential-sol-real}
\end{align}

\begin{figure}[h!!]
\begin{center}
\includegraphics[width=0.32\textwidth]{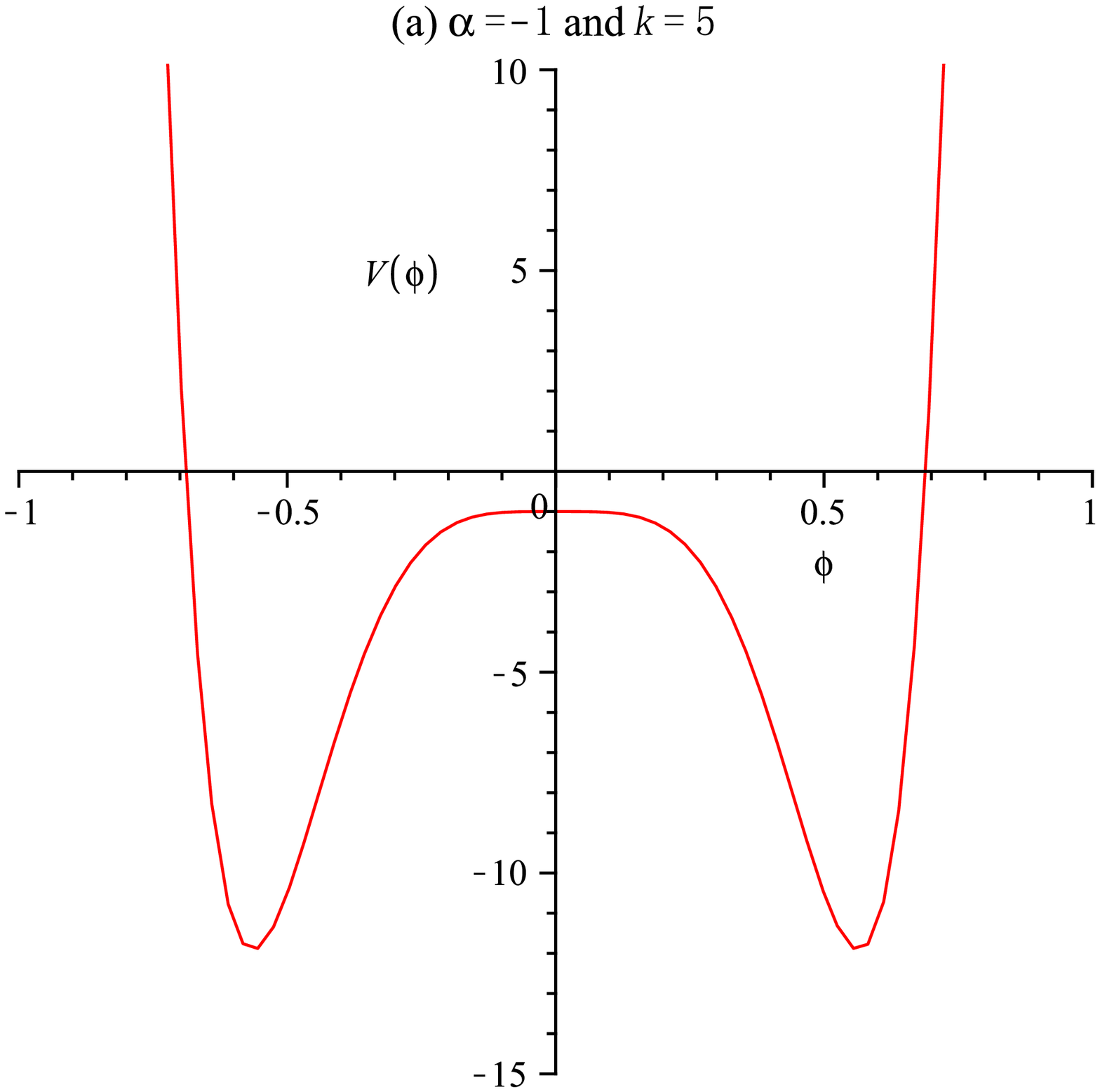}
\includegraphics[width=0.32\textwidth]{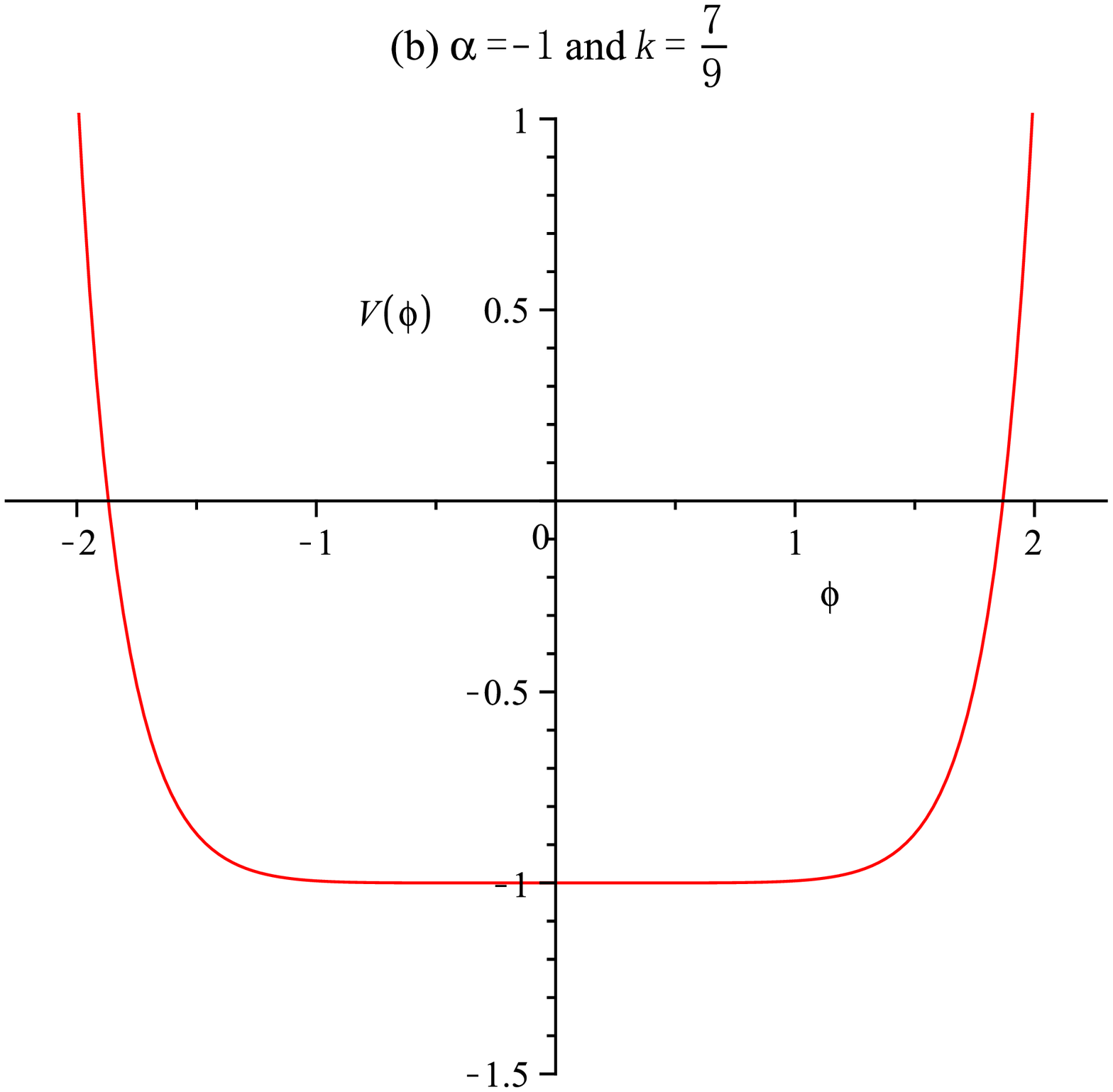}
\includegraphics[width=0.32\textwidth]{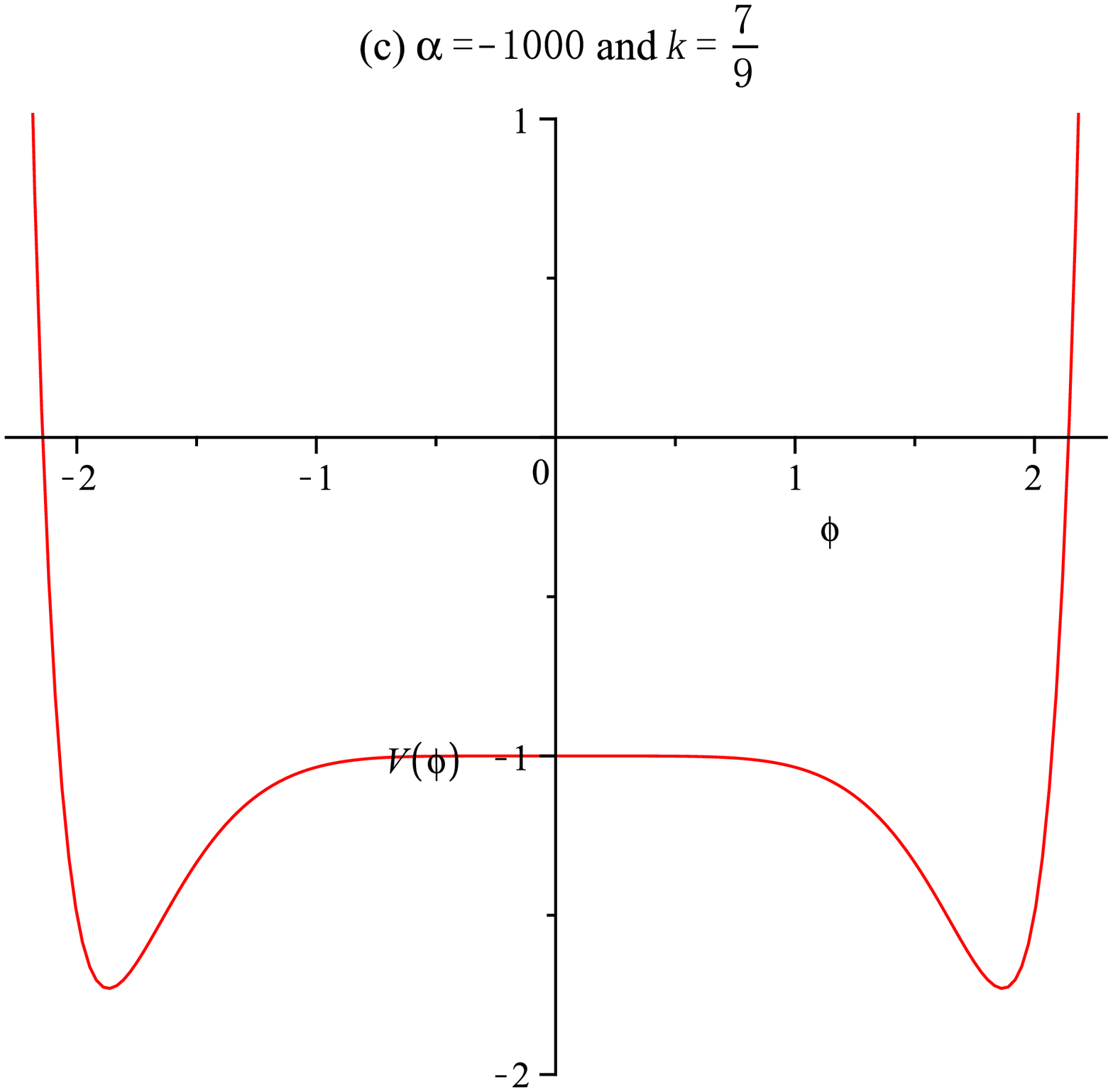}
\caption{ Self-coupling potential $V(\phi)$ versus $\phi$ with $q=2$ and $\ell=1$.}\label{potential-fig}
\end{center}
\end{figure}

On the other hand, when $\phi^2\rightarrow0$, the leading terms in the power series expansion of $V(\phi)$ behave as
\begin{align*}
  V(\phi)=-\frac{1}{\ell^2}+\frac{1}{512}\left(\frac{1}{\ell^2}+\frac{\alpha}{48}\right)\phi^6
  +\frac{(2k-1)2^{(k-1)}\pi{q}^{2k}}{2^{\frac{6k}{(2k-1)}}}\phi^{\frac{4k}{(2k-1)}}\nonumber\\
  \times\left(-\frac{k}{8(k-1)(4k-1)}\phi^2+\frac{k}{32(2k-1)}\phi^4+O(\phi^6)\right),
\end{align*}
where we have inserted Eq.(\ref{k-value}). One can find some discussions:
\begin{itemize}
\item If $k\in(1,+\infty)$, hence $\frac{4k}{(2k-1)}<4$. When $\phi$ is small, the main contribution of $V(\phi)$ is from $-\frac{k(2k-1)}{8(k-1)(4k-1)}\frac{2^{(k-1)}\pi{q}^{2k}}{2^{\frac{6k}{(2k-1)}}}\phi^{\frac{4k}{(2k-1)}+2}$
    term whose coefficient is negative. Thus $V(\phi)$ will possess a maximum at $\phi=0$
    and two minimum at some non-vanishing constant scalar field. This is more easily
seen in the left plot of Fig.(\ref{potential-fig}).

\item If $k\in(\frac{1}{2},1)$, then $\frac{4k}{(2k-1)}>4$. When $\phi$ is small, the main contribution of $V(\phi)$ is from $\frac{1}{512}\left(\frac{1}{\ell^2}+\frac{\alpha}{48}\right)\phi^6$ term whose
    coefficient determines the behavior of potential. If, in addition,
$\left(\frac{1}{\ell^2}+\frac{\alpha}{48}\right)>0$, $V(\phi)$ will only possess a minimum at $\phi=0$.
This can be seen in the middle plot of Fig.(\ref{potential-fig}). If $\left(\frac{1}{\ell^2}
+\frac{\alpha}{48}\right)\leq0$, $V(\phi)$ will also possess a maximum at $\phi=0$ and two minimum.
One can see it in the right plot of Fig.(\ref{potential-fig}).

\end{itemize}

For all above cases, one can find that the system is always stable against small perturbations in $\phi$.

\subsection{Geometric properties and horizon structure}
For this charged hairy black hole, the mass can be calculated simply
by adopting the Brown-York method \cite{Brown:1992br}. The quasilocal mass $m(r)$ at a $r$
takes the following form \cite{Brown:1994gs,Creighton:1995au,Brown:1992br}
\begin{eqnarray}
m(r)=2\sqrt{f(r)}(\sqrt{f_0(r)}-\sqrt{f(r)}).\label{3b}
\end{eqnarray}
As a result, the mass can be obtained as
\begin{eqnarray}
M\equiv \lim_{r\rightarrow\infty}m(r)=-\frac{\alpha B^2}{16},\label{4b}
\end{eqnarray}
using which, we can rewrite the horizon function as
\begin{align}
  f(r)=\frac{r^2}{\ell^2}-\left(1+\frac{2B}{3r}\right)M
  -\frac{X(2k-1)\left(r+\frac{2Bk}{(4k-1)}\right)}{r^{\frac{1}{(2k-1)}}}.\label{metric-sol-real}
\end{align}
where $X=\frac{2^k(2k-1)\pi{Q}^{2k}}{2(k-1)}$.

In order to have a further understanding of the solutions, we calculate some geometric quantities.
Firstly, the Ricci scalar reads as
\begin{align*}
  R=-\frac{6}{\ell^2}+\frac{2(k-1)X}{(2k-1)}\left((4k-3)-\frac{2Bk}{(4k-1)r}\right)r^{\frac{-2k}{(2k-1)}},
\end{align*}
which shows that the solution has an essential singularity at $r=0$ whenever $Q\neq0$.
Higher order curvature invariants such as $R_{\mu\nu}R^{\mu\nu}$ and $R_{\mu\nu\rho\sigma}R^{\mu\nu\rho\sigma}$
both have much complicated expressions. One can also find the further behavior as $O(\frac{M}{r})$, hence the solution is also singular at $r=0$ whenever $M\neq0$.
As we are interested in the black holes, the solutions need to contain a event horizon
to surround the singularity.

On the other hand, in order to indicate that the metric is non-conformally flat,
we can find some non-vanishing components of the Cotton tensor, e.g.
\begin{align*}
  C_{\theta\theta\,r}=\frac{MB}{r^2}+\frac{k
  \bigg((1-k)+\frac{Bk}{r}\bigg)X}{(2k-1)^2r^{\frac{1}{(2k-1)}}},
\end{align*}
which does not vanishes whenever $B\neq0$, $M\neq0$ or $Q\neq0$.

When $Q=0$, the solutions reduce to the uncharged scalar solutions, whose horizon structure
is given in \cite{Xu:2013nia}. In what follows, we will only focus on the charged case.
However, the expression of $f(r)$ with $Q\neq0$ is complicated so that we can not solve
directly the black hole horizons. Thus we focus on the extreme charged black hole firstly.
As the Hawking temperature is
\begin{align}
  T&=\frac{f'(r_+)}{4\pi}=\frac{1}{2\pi}\left[\frac{r_{+}}{\ell^2}+\frac{BM}{3r_{+}^2}
  -\frac{X}{(2k-1)}\frac{\left((k-1)-\frac{Bk}{(4k-1)r_{+}}\right)}{r_{+}^{\frac{1}{(2k-1)}}}\right].
\end{align}
Now we can evaluate the extreme charged black hole $T=0$. Then the mass of extremal black hole is
\begin{align}
  M_{ex}=-\frac{3(4k-1)kr_{ex}^3}{(k-1)\bigg(3(4k-1)r_{ex}+4Bk\bigg)\ell^2},
\label{M-ex}
\end{align}
where the extreme black hole horizon $r_{ex}(X)$ is the biggest root of the following equation
\begin{align}
  X=\frac{3(2k-1)(4k-1)r_{ex}^{\frac{(4k-1)}{(2k-1)}}}{(k-1)\bigg(3(4k-1)r_{ex}+4Bk\bigg)\ell^2}.
\label{r-ex}
\end{align}

As it is still difficult to find analytical result for $r_{ex}(X)$ and $M_{ex}$,
we can consider another way to know more about them. From Eq.(\ref{M-ex}) and Eq.(\ref{r-ex}),
we can rewrite the the extreme black hole horizon as
\begin{align}
  r_{ex}(M_{ex},X)=\left(-\frac{Xk}{(2k-1)M_{ex}}\right)^{\frac{(2k-1)}{2(1-k)}},
\label{r-ex1}
\end{align}
while the condition for extreme charged black hole, i.e. $M_{ex}$, being the biggest root of the following equation
\begin{align}
 M_{ex}=-\frac{3(4k-1)kr_{ex}(M_{ex},X)^3}{(k-1)\bigg(3r_{ex}(M_{ex},X)(4k-1)+4Bk\bigg)\ell^2}.
\label{M-ex1}
\end{align}
As $r_{ex}$ is real and positive, Eq.(\ref{r-ex1}) reveals the classification of the
horizon structure of this solutions, which are divided into the following cases:
\begin{itemize}
\item for $k\in(\frac{1}{2},1)$: one can find $X<0$, thus $M_{ex}>0$:
  \begin{enumerate}
    \item When $M<M_{ex}$, the solutions have no horizon,
    thus there is a bare singularity in the origin of the spacetime which is physically unaccepted;
  \item When $M=M_{ex}$, the solutions have a single horizon $r_{ex}$,
  which corresponds to the extreme black hole;
  \item When $M>M_{ex}$, the solutions have two horizons $r_{E}$ and $r_{C}$,
  which corresponds to the non-extreme black hole.
  \end{enumerate}

\item for $k\in(1,+\infty)$: one can find $X>0$, thus $M_{ex}<0$:
  \begin{enumerate}
  \item When $M<M_{ex}$, the solutions have no horizon which leads to a bare singularity
  in the origin of the spacetime, thus it is physical unaccepted;
  \item When $M=M_{ex}$, the solutions have a single  horizon $r_{ex}$. This corresponds to the extreme black hole;
  \item When $M_{ex}<M<0$, the solutions have two horizons $r_{E}$ and $r_{C}$. This is the non-extreme black hole;
  \item When $M\geq0$, there is no another extreme black hole case for solutions
  with positive mass, thus these solutions can only have either zero or one horizon.
  Besides, as $f(0)f(+\infty)<0$,  $f(r)$ must come across zero once at least,
  thus the solutions are the black holes with a single horizon.
  \end{enumerate}
\end{itemize}

\begin{figure}[h!]
\begin{center}
\includegraphics[width=0.45\textwidth]{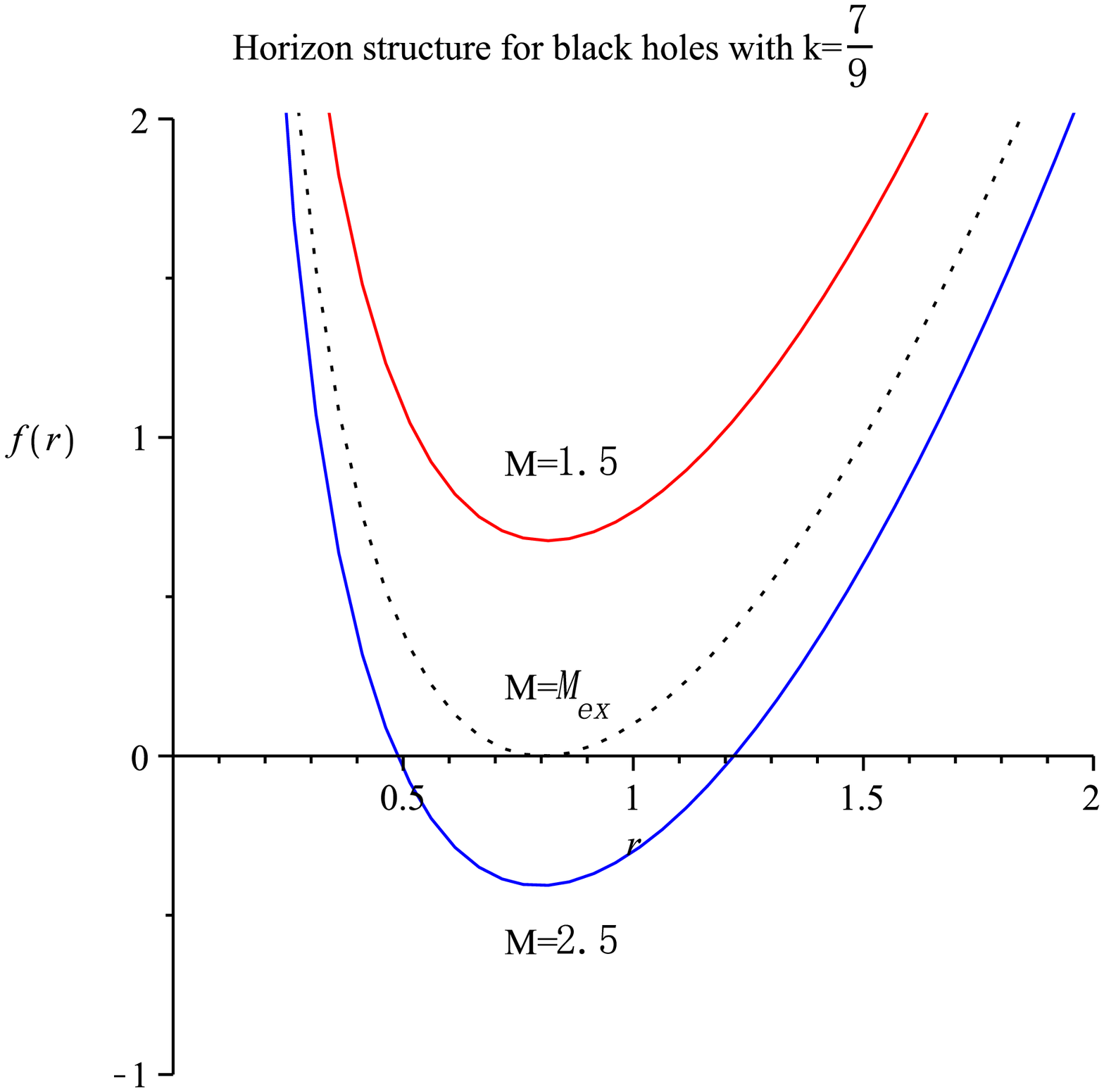}
\includegraphics[width=0.45\textwidth]{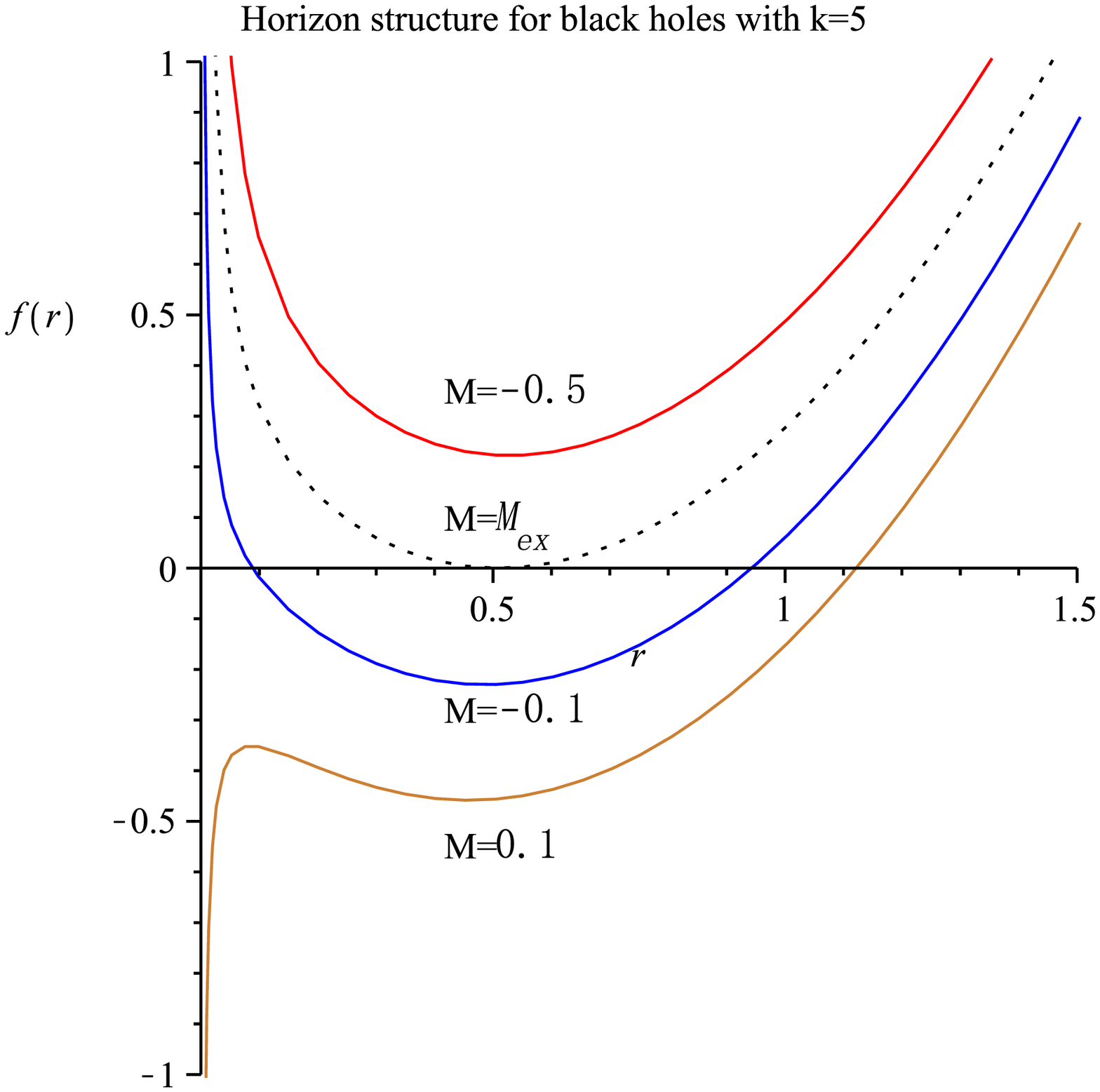}
\caption{The horizon structure with $Q=0.5,B=0.1,\ell=1$. On both plots, the mass decreases from top to bottom.
The dashed curve corresponds to the extreme black holes. The mass for the extreme black hole
with $k=\frac{7}{9}$ on the left is $M_{ex}\simeq2.1242$, while it for $k=5$ on the right is $M_{ex}\simeq-0.3030$.}
\label{horizon-fig}
\end{center}
\end{figure}

In Fig.(\ref{horizon-fig}), we plot the horizon function $f(r)$ with $Q=0.5,B=0.1,\ell=1$
and different values of parameters $M,k$, in order to have a more clear understanding
of the above horizon structure.

Totally, in order to avoid the appearance of a naked singularity,
a horizon in the spacetime is needed at least. In this sense,
we see that the physically acceptable region for mass $M$ is $M\geq\,M_{ex}$ for any values of parameter $k$. Note for the cases $k>1$, the lower bound $M_{ex}$ is negative, which shows that there exist scalar black holes with negative mass in three dimensions. Actually, the negative mass is acceptable, as it is well-known that the mass
for some kinds of black holes is bounded from
below by a negative value \cite{Martinez:2005di,Vanzo:1997gw,Mann:1997jb,Brill:1997mf,Belletete:2013nqa,Mbarek:2014ppa},
especially for the four dimensional charged scalar black holes \cite{Martinez:2005di}.
Obviously, the three dimensional charged scalar black holes with $k\in(1,+\infty)$
in EPM theory present in this paper share the same feature.

For the analytical results, we can give some detailed discussions about the case with $B=0$,
which corresponds to the black holes without scalar hair in EPM theory.
For this simplified case, Eq.(\ref{M-ex}) and Eq.(\ref{r-ex}) reduce to
\begin{align*}
  M_{ex}=-\frac{kr_{ex}^2}{(k-1)\ell^2},\quad\,X=\frac{(2k-1)r_{ex}^{\frac{2k}{(2k-1)}}}{(k-1)\ell^2},
\end{align*}
which lead to the condition and horizon for extreme charged black hole
\begin{align}
  r_{ex}=(2Q^2)^{\frac{(2k-1)}{2}}\left(\frac{\pi}{2}\ell^2\right)^{\frac{(2k-1)}{2k}},\quad
  \,M_{ex}=-\frac{k}{(k-1)\ell^2}(2Q^2)^{(2k-1)}\left(\frac{\pi}{2}\ell^2\right)^{\frac{(2k-1)}{k}}.
\label{M-ex1}
\end{align}
Consider the parameter $k$: for $k\in(\frac{1}{2},1)$, $M_{ex}>0$; while for $k\in(1,+\infty)$, we get $M_{ex}<0$.
The above discussion about the horizon structure still works.
Hence we can see that the scalar field does not affect the horizon structure,
which is consistent with the $k=1$ case shown in \cite{Xu:2013nia,Xu:2014uha}.
The effect of scalar field is only related to the value of horizons and $M_{ex}$.
This can be seen from comparing with the horizon structure of black
holes with $B=0.1$ shown in Fig(\ref{horizon-fig}). Let us consider the
black holes with $B=0,Q=0.5,\ell=1$, then Eq.(\ref{M-ex1}) shows that the
mass for extreme black hole with $k=\frac{7}{9}$ is $M_{ex}\simeq2.1606$,
while it for $k=5$ is $M_{ex}\simeq-0.2873$.

\section{Null geodesics}\label{3s}
Let us consider the geodesic equations for uncharged test particles around the
solution with scalar hair in EPM Theory. Since this spacetime has two
Killing vectors $\partial_t$ and $\partial_\theta$, there are two constants of motions, i.e.
\begin{align}
  &E=f(r)\frac{\mathrm{d} t}{\mathrm{d} \lambda},\label{E}\\
  &L=r^2\frac{\mathrm{d} \theta}{\mathrm{d} \lambda}\label{L},
\end{align}
where $\lambda$ is the affine parameter along the geodesics.
The geodesic equation can be derived from the Lagrangian for a test particle
\begin{align*}
  f(r)\left(\frac{\mathrm{d} t}{\mathrm{d} \lambda}\right)^2-\frac{1}{f(r)}\left(\frac{\mathrm{d} r}{\mathrm{d} \lambda}\right)^2-r^2\left(\frac{\mathrm{d} \theta}{\mathrm{d} \lambda}\right)=-m^2,
\end{align*}
where $m=0$ corresponding to null geodesics and $m=1$ corresponding to
time-like geodesics (without loss of generality). Inserting Eq.(\ref{E}) and Eq.(\ref{L})
into the above equation, one can obtain the effective equation
\begin{align}
  \frac{1}{2}\left(\frac{\mathrm{d} r}{\mathrm{d} \lambda}\right)^2+V_{eff}(r)=0,\label{eff-eq}
\end{align}
where $V_{eff}(r)$ is the effective potential and takes the form as
\begin{align}
  V_{eff}(r)=\frac{1}{2}\bigg[f(r)\left(\frac{L^2}{r^2}+m^2\right)-E^2\bigg].
\label{eff-potential}
\end{align}
Then from Eq.(\ref{L}) and Eq.(\ref{eff-eq}), we get the orbit equation
\begin{align}
  \left(\frac{\mathrm{d} r}{\mathrm{d} \theta}\right)^2=-\frac{2r^4}{L^2}V_{eff}(r)
  \label{orbit}
\end{align}

As we consider the null geodesics, i.e. the geodesics for a photon. The effective potential (\ref{eff-potential}) reduces to
\begin{align}
  V_{eff}(r)=\frac{1}{2}\bigg[f(r)\left(\frac{L^2}{r^2}\right)-E^2\bigg].
\label{eff-potential1}
\end{align}
In the following subsections, we focus on the orbit equation (\ref{orbit})
and effective potential (\ref{eff-potential1}) to classify all possible geodesic motions.

\subsection{Radial geodesics where $L=0$}
Firstly we examine the radial geodesics where $L=0$.
The corresponding effective potential Eq.(\ref{eff-potential1}) further reduces to
\begin{align*}
  V_{eff}(r)=-\frac{E^2}{2}.
\end{align*}
Obviously, the behavior of these geodesics do not depend on the electric charge $Q_e$
and mass $M$ of the black hole. As $E$ is a constant, this resembles the geodesic motion of a free photon.

Combining Eq.(\ref{E}) and Eq.(\ref{eff-eq}), we can get
\begin{align}
 \frac{\mathrm{d} t}{\mathrm{d} r}=\pm\frac{1}{f(r)}.
 \label{time}
\end{align}
Here we are only interested in the geodesics of the black holes, then one can rewrite the metric function as
\begin{align}
  f(r)=(r-r_{E})(r-r_{C})F(r)
\end{align}
with $r_{E}$ being the Event horizon and $r_{C}$ being the Cauchy horizon of the black holes,
respectively. However, it is difficult to find the form of $F(r)$, but we still know that $F(r)$
can have either no real roots or negative roots. For non-extreme charged black hole
with two horizons, $r_{E}$ and $r_{C}$ are both positive. For charged black hole with single horizon, only $r_{E}$ is real and positive. For extreme charged black hole, $r_{E}=r_{C}$ is positive.

Rewriting the right side of Eq.(\ref{time}), it is equal to
\begin{align*}
  \frac{1}{f(r)}&=\frac{1}{(r_{E}-r_{C})}\bigg[\frac{1}{F(r_{E})}\bigg(\frac{1}{(r-r_{E})}
  -\frac{G(r,r_{E})}{F(r)} \bigg)\nonumber\\
  &-\frac{1}{F(r_{C})}\bigg(\frac{1}{(r-r_{C})}-\frac{G(r,r_{C})}{F(r)} \bigg)\bigg]
\end{align*}
for the non-extreme black hole, where $G(r,r_{i})=\frac{F(r)-F(r_i)}{(r-r_i)}$ with $i$ being $(E,C)$.
Note $G(r_{i},r_{i})=F'(r_{i})$ is finite. After integrating Eq.(\ref{time}), we find
\begin{align*}
  t&=\pm\frac{1}{(r_{E}-r_{C})}\bigg[\frac{1}{F(r_{E})}\bigg(\ln(r-r_{E})-\int\frac{G(r,r_{E})}{F(r)}\mathrm{d}r \bigg)\nonumber\\
  &-\frac{1}{F(r_{C})}\bigg(\ln(r-r_{C})-\int\frac{G(r,r_{C})}{F(r)}\mathrm{d}r \bigg)\bigg].
\end{align*}
Similarly, for the extreme black hole case and the black hole case with a single horizon, it leads to
\begin{align*}
   t=\pm\frac{1}{F(r_{E})}\bigg(\frac{1}{(r_{E}-r)}-\int\frac{G(r,r_{E})}{F(r)(r-r_{E})}\mathrm{d}r \bigg).
\end{align*}
For all cases, the sign $``+"$ denotes the out going null rays and the sign $``-"$ denotes the ingoing null ray.
Consider the ingoing null rays, when $r\rightarrow\,r_{E}$, the coordinate time $t\rightarrow+\infty$ for the black holes.

On the other hand, the geodesic equation Eq.(\ref{eff-eq}) can be integrated to give
\begin{align*}
  r(\lambda)=\pm\,E \lambda,
\end{align*}
From which, we find when $r\rightarrow\,r_{E}$ (in-going case), $\lambda$ has a finite
value $\frac{r_{E}}{E}$. Hence one can see that a photon without angular momentum
arrives the horizons in its own finite proper time,
while it is an infinite coordinate time.

\subsection{Radial geodesics where $L\neq0$}
Then we consider the radial geodesics where $L\neq0$. From Eq.(\ref{eff-potential1}), the effective potential can be
obtained as
\begin{align}
  V_{eff}(r)=\frac{L^2}{2r^2}\bigg[\left(\frac{1}{\ell^2}-\Xi^2\right)r^2
  -\left(1+\frac{2B}{3r}\right)M-\frac{X(2k-1)\left(r+\frac{2Bk}{(4k-1)}\right)}{r^{\frac{1}{(2k-1)}}}\bigg],
\end{align}
where $\Xi=\frac{E}{L}$.
When the minimum value of $V_{eff}$ is zero, the geodesic reaches to the circular motion. This leads to the critical condition
\begin{align}
  \Xi_{cri}=\sqrt{\frac{1}{\ell^2}+\frac{(k-1)M}{r_{cri}^2}\left(\frac{1}{k}+\frac{4B}{3(4k-1)r_{cri}}\right)}
\label{Xi-cri}
\end{align}
and the radius of the circular motion
\begin{align}
  r_{cri}(M,X)=\left(-\frac{Xk}{(2k-1)M}\right)^{\frac{2k-1}{2(1-k)}}
\label{r-cri}
\end{align}

\begin{figure}[h!]
\begin{center}
\includegraphics[width=0.32\textwidth]{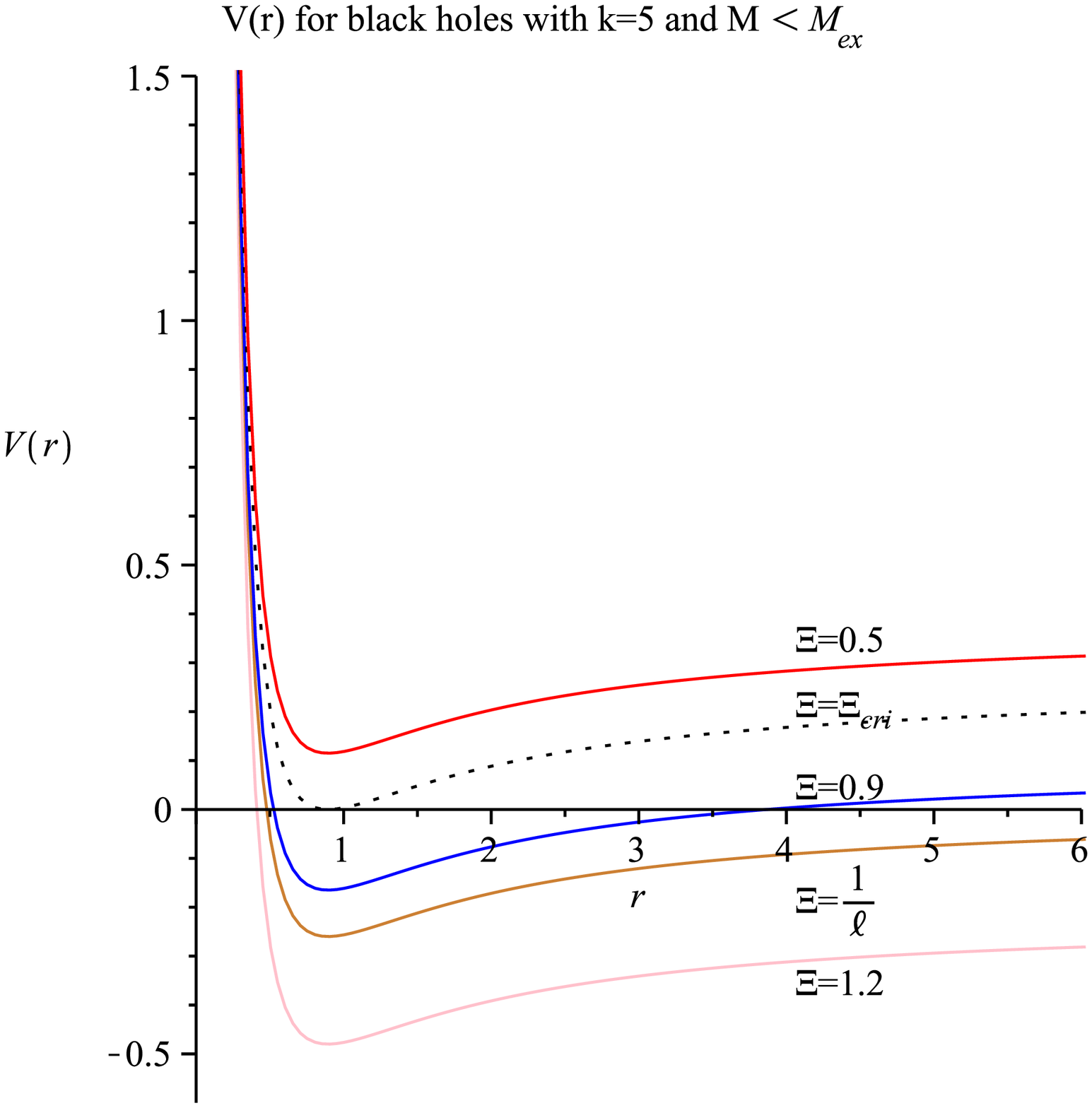}
\includegraphics[width=0.32\textwidth]{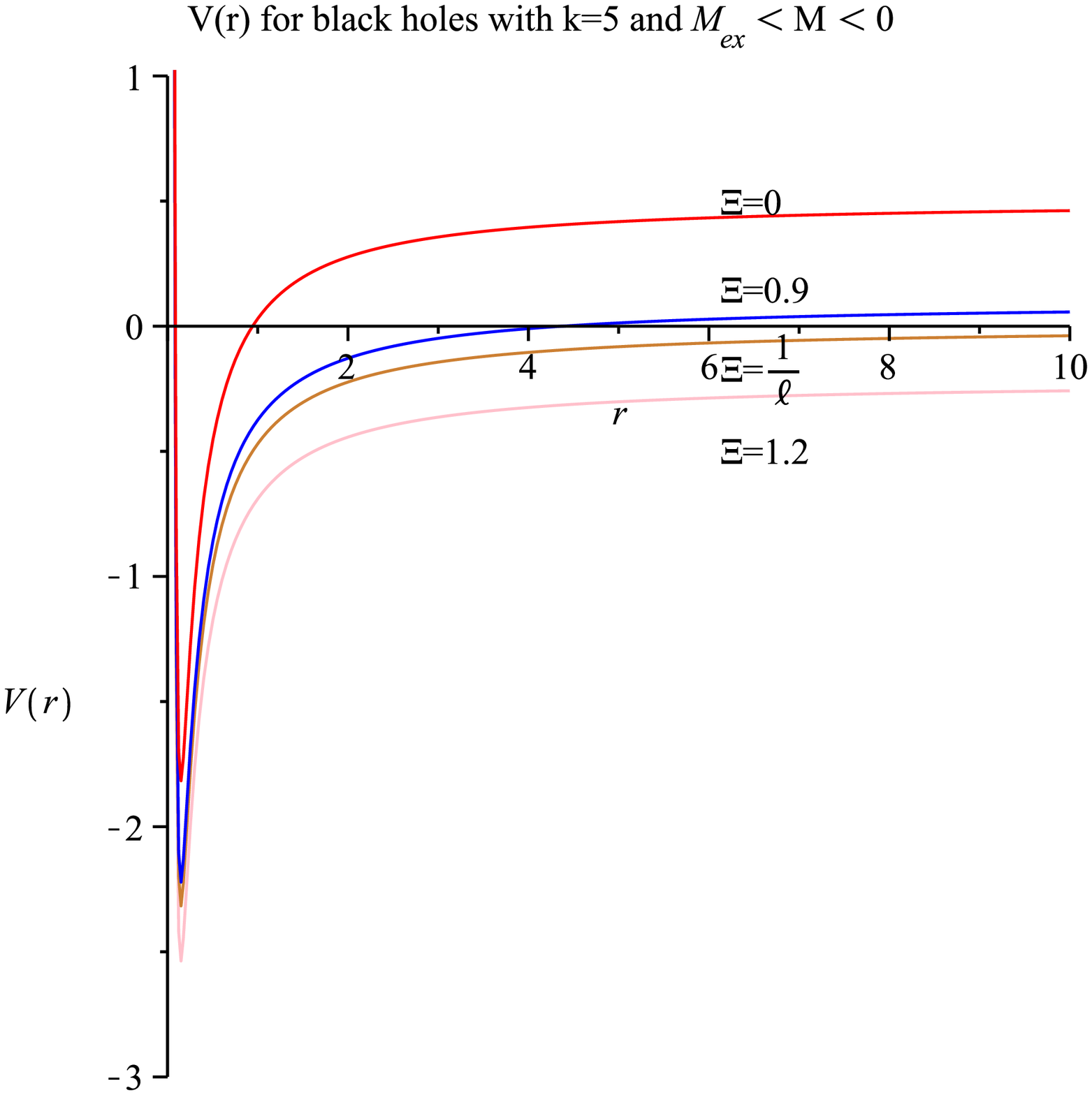}
\includegraphics[width=0.32\textwidth]{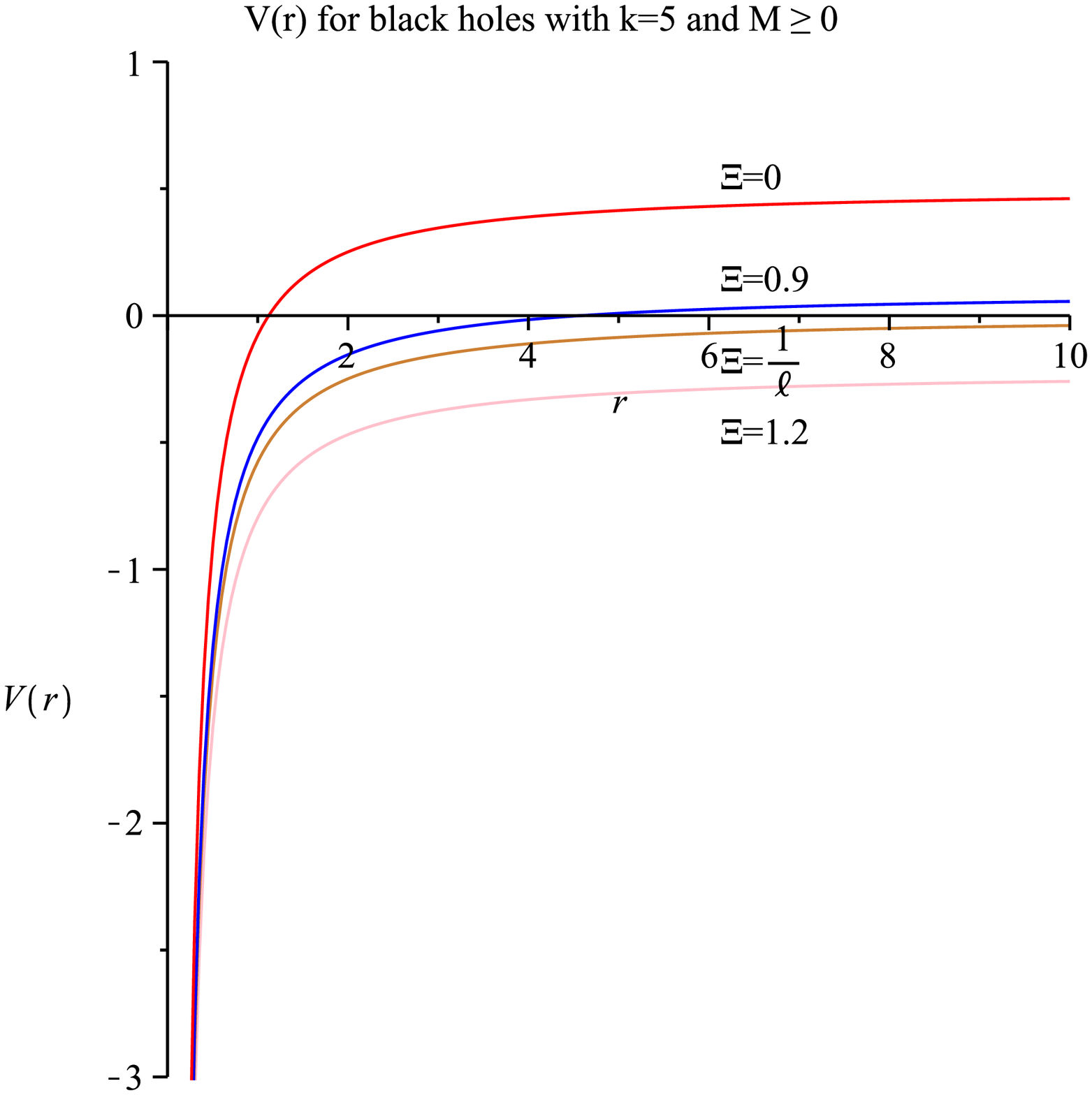}
\caption{The effective potential $V(r)$ with $k=5,Q=0.5,B=0.1,\ell=1$. For this case,
the mass of extreme black holes $M_{ex}\simeq-0.3030$. From left plot to right,
the mass for solutions are $M=-0.5,M=-0.1$ and $M=0.1$, respectively.
On all plots, the constant of motions $\Xi$ decreases from top to bottom.
The dashed curve corresponds to the case for circular motion.}
\label{geodesics2}
\end{center}
\end{figure}

As $r_{cri}$ is real and positive, Eq.(\ref{r-cri}) shows that $r_{cri}$ can only exist in two cases:
(1) for $k\in(\frac{1}{2},1)$, we get $X<0$, thus $M>0$ is needed;
(2) for $k\in(1,+\infty)$, we find $X>0$, thus $M<0$ is needed.
For other cases, $r_{cri}$ is imaginary, hence the circular motion can not exist.
However, even for the above two cases, one can find $(k-1)M<0$, thus Eq.(\ref{Xi-cri})
leads to another constraint on $M$, in order to keep $\Xi_{cri}$ real. For this one,
we can firstly focus on the critical case, i.e. $\Xi_{cri}=0$ with the effective potential
$V(r)=\frac{L^2}{2r^2}f(r)$. Then the vanishing minimum of this effective potential
is equal to that of horizon function $f(r)$, which immediately leads to the critical
condition $M=M_{ex}$ and the radius of circular motion $r_{cri}=r_{ex}$. However,
when $\Xi_{cri}>0$, it is difficult to find the physical constraint of $M$,
because $r_{cri}$ is not equal to $r_{ex}$ in this case. But the discussion about
the effective potential in what follows will show that the constraint for real $\Xi_{cri}$
is $M\leq\,M_{ex}$ for $k\in(\frac{1}{2},+\infty)$. (One can also see Fig.\ref{geodesics1} and Fig.\ref{geodesics2}.)

Putting the above discussions together, we can obtain
\begin{itemize}
   \item for $k\in(\frac{1}{2},1)$: as $M_{ex}>0$, one can get that only the geodesics in the spacetime
   with $0<M\leq\,M_{ex}$ can contain the circular motion with its radius $r=r_{cri}$;
   \item for $k\in(1,+\infty)$: as $M_{ex}<0$, only the geodesics in the spacetime with $M\leq\,M_{ex}(<0)$
   can contain the circular motion with the radius $r=r_{cri}$.
\end{itemize}

These will classify the discussion of the geodesics.
One can find that the geodesics are divided into several cases according to the above constraint
and the horizon structure. This can also be seen from the behavior of the effective potntial,
which is shown in Fig.\ref{geodesics1} and Fig.\ref{geodesics2}. Actually, there are five models in these spacetime:

\begin{figure}[h!]
\begin{center}
\includegraphics[width=0.32\textwidth]{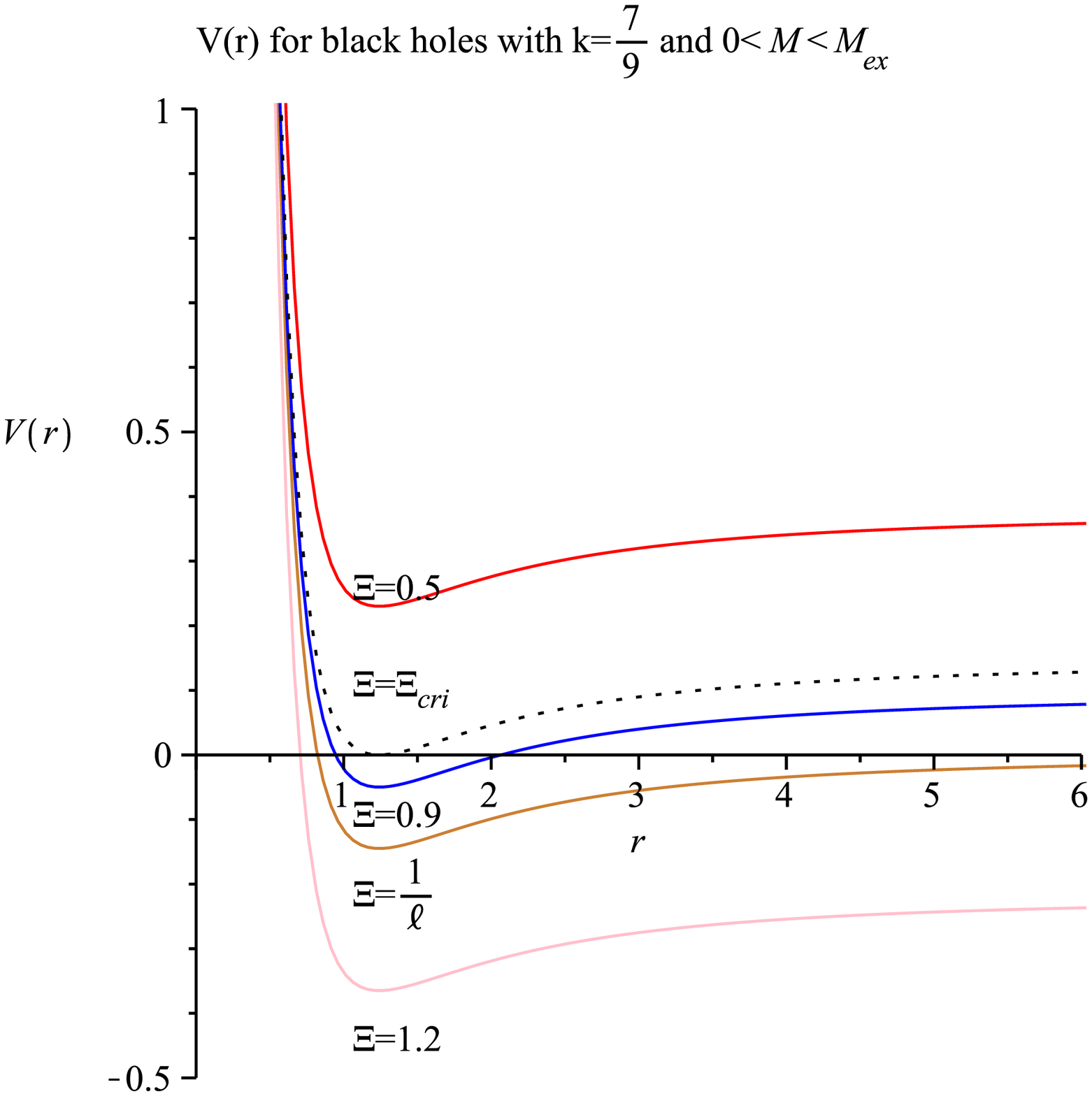}
\includegraphics[width=0.32\textwidth]{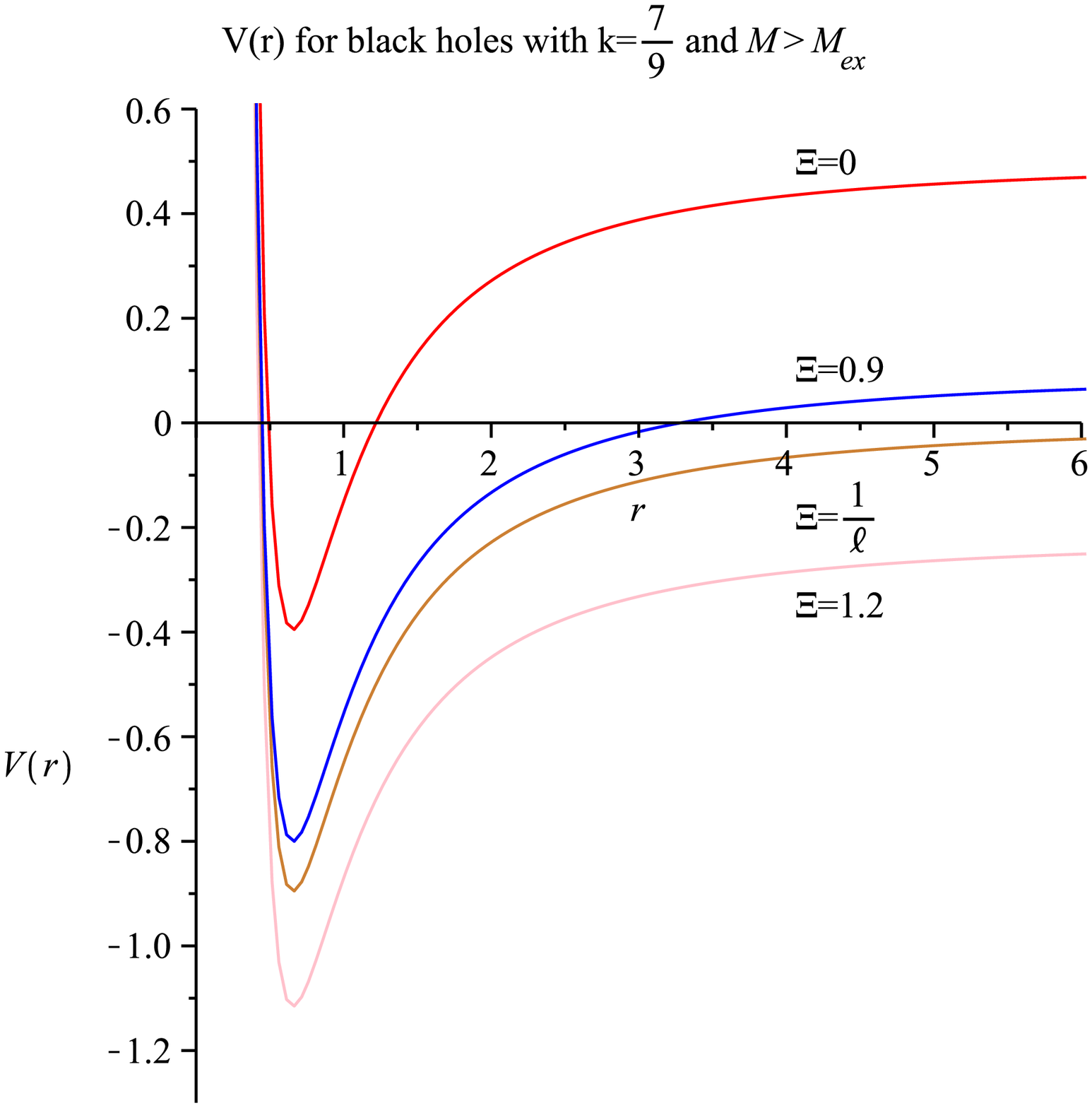}
\includegraphics[width=0.32\textwidth]{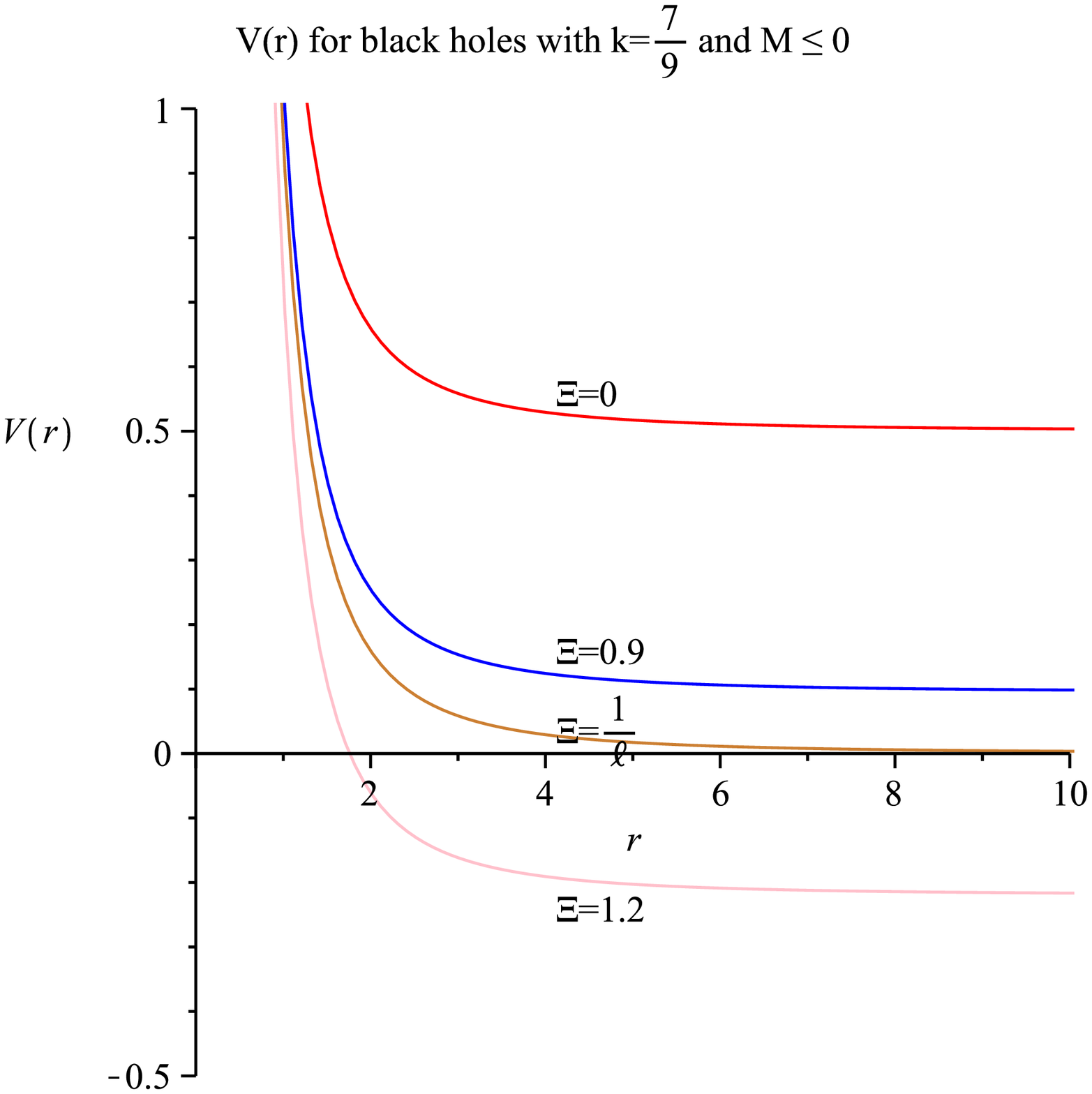}
\caption{The effective potential $V(r)$ with $k=\frac{7}{9},Q=0.5,B=0.1,\ell=1$. For this case,
the mass of extreme black holes $M_{ex}\simeq2.1242$. From left plot to right,
the mass for solutions are $M=1.5,M=2.5$ and $M=-0.5$, respectively.
On all plots, the constant of motions $\Xi$ decreases from top to bottom.
The dashed curve corresponds to the case with circular motion.}
\label{geodesics1}
\end{center}
\end{figure}

\begin{itemize}
\item Model I: this corresponds to the cases with $\frac{1}{2}<k<1,0<M<\,M_{ex}$ and $k>1,M<\,M_{ex}$. These corresponding metric do not have a horizon. For the effective potential, one can look at the left plots of Fig.\ref{geodesics1} and Fig.\ref{geodesics2}, respectively. When the constant of motions $\Xi$ changes, all the geodesic motions of a photon can be categorized by four subcases:

  \begin{enumerate}
    \item when $0\leq\Xi<\Xi_{cri}$, the effective potential is always positive in the whole spacetime. From the orbit equation Eq.(\ref{orbit}), one can only get a imaginary orbit equation, thus there is no allowed motion. Actually, the orbits are allowed only when $\Xi\geq\Xi_{cri}$ as shown later.
    \item when $\Xi=\Xi_{cri}$, the photon can only stay at the circular motion with the radius $r_{cri}$. For example, the critical quantities in the left plots of Fig.\ref{geodesics1} and Fig.\ref{geodesics2} are $\Xi_{cri}\simeq0.8427,r_{cri}\simeq1.2398$ and $\Xi_{cri}\simeq0.6929,r_{cri}\simeq0.8942$, respectively.
    \item when $\Xi_{cri}<\Xi<\frac{1}{\ell}$, the photon has the elliptic motions. Namely, all orbits are bounded between aphelion and perihelion. For this subcase, we can only get the numerical results; consider the cases with $\Xi=0.9$ in the left plots of Fig.\ref{geodesics1} and Fig.\ref{geodesics2}, the aphelion and perihelion are $r_{ap}\simeq2.0658,r_{ph}\simeq0.9439$ and $r_{ap}\simeq3.8709,r_{ph}\simeq0.5239$,respectively.
    \item when $\Xi\geq\frac{1}{\ell}$, the geodesic motions are unbounded spiral motions at a large scale. . When $\Xi>\frac{1}{\ell}$, the orbit equation Eq.(\ref{orbit}) is not integrable, thus the numerical analysis is a useful tool for those geodesic motions. Note that there also exists perihelion but we cannot obtain it analytically. However, when $\Xi=\frac{1}{\ell}$, Eq.(\ref{orbit}) becomes integrable, which is shown in detail here. For simplicity, we consider the case with $k=\frac{7}{9},Q=\frac{1}{2},B=0$, then the explicit form of the integrable orbits corresponding to $\Xi=\frac{1}{\ell}$ is
        \begin{align}
          e^{\frac{2}{5}\sqrt{M}\theta}=12\sqrt{M}r^{\frac{2}{5}}
          +\sqrt{2\left(72Mr^{\frac{4}{5}}-25\times{2}^{\frac{2}{9}}\pi\right)},
        \label{analytical}
        \end{align}
        which can be simplified to
        \begin{align}
          r=\left[\frac{1}{24\sqrt{M}}\bigg(e^{\frac{2}{5}\sqrt{M}\theta}+50\times{2}^{\frac{2}{9}}\pi\,e^{-\frac{2}{5}\sqrt{M}\theta}\bigg)\right]^{\frac{5}{2}}.
        \label{analytical2}
        \end{align}
        The perihelion of this analytical orbit in Eq.(\ref{analytical}) is trivially obtained as
        \begin{align}
          r_{ph}=\left[\frac{1}{24\sqrt{M}}\bigg(1+50\times{2}^{\frac{2}{9}}\pi\bigg)\right]^{\frac{5}{2}}.
        \label{analytical-ph}
        \end{align}
  \end{enumerate}

\item Model II: this corresponds to the cases with $\frac{1}{2}<k<+\infty,M=M_{ex}$. These solution are the extreme black holes. For this cases, the effective potential has the same shape with the left plots of Fig.\ref{geodesics1} and Fig.\ref{geodesics2}. The only difference is $\Xi_{cri}$ will reach to zero. As a result, the circular motion for this extremal charged EPM scalar black hole is the stopped motion at the degenerated horizon $r_{cri}=r_{ex}$, which means eventually that this circular motion is not allowed. Then when the constant of motions $\Xi$ changes, all the geodesic motions of a photon can be categorized by three subcases:

  \begin{enumerate}
    \item when $\Xi=0$, there is no allowed motion, as the circular motion is not physical.
    \item when $0<\Xi<\frac{1}{\ell}$, the photon has the similar elliptic motions as that of Model I.
    \item when $\Xi\geq\frac{1}{\ell}$, the geodesic motions of the photon are also unbounded spiral motions. We only consider the integrable case with $\Xi=\frac{1}{\ell}$ here. In additional with $k=\frac{7}{9},Q=\frac{1}{2},B=0$, the orbits can be trivially obtained from Eq.(\ref{analytical2})
        \begin{align*}
          r=\left[\frac{1}{24\sqrt{M_{ex}}}\bigg(e^{\frac{2}{5}\sqrt{M_{ex}}\theta}+50\times{2}^{\frac{2}{9}}\pi\,e^{-\frac{2}{5}\sqrt{M_{ex}}\theta}\bigg)\right]^{\frac{5}{2}}
        \end{align*}
        with the perihelion obtained from Eq.(\ref{analytical-ph})
        \begin{align*}
          r_{ph}^{ex}=\left[\frac{1}{24\sqrt{M_{ex}}}\bigg(1+50\times{2}^{\frac{2}{9}}\pi\bigg)\right]^{\frac{5}{2}}.
        \end{align*}
  \end{enumerate}

\item Model III: this corresponds to the cases with $\frac{1}{2}<k<1,M>\,M_{ex}$ and $k>1,M_{ex}<M<0$.  These solutions have two horizons. For the effective potential, one can focus on the middle plots of Fig.\ref{geodesics1} and Fig.\ref{geodesics2}, respectively. From the figures, we can get that the subcase without allowed motion does not exist, i.e. all subcases have allowed motions. When the constant of motions $\Xi$ changes, all the geodesic motions of a photon can be categorized by two subcases:

  \begin{enumerate}
    \item when $0\leq\Xi<\frac{1}{\ell}$, the geodesic motions are the elliptic motions. Consider the subcases with $\Xi=0.9$ in the middle plots of Fig.\ref{geodesics1} and Fig.\ref{geodesics2}, the aphelion and perihelion are $r_{ap}\simeq3.2832,r_{ph}\simeq0.4446$ and $r_{ap}\simeq4.3720,r_{ph}\simeq0.0849$,respectively.
    \item when $\Xi\geq\frac{1}{\ell}$, the photon stays at the unbounded spiral motions. The corresponding analytical orbits for $\Xi=\frac{1}{\ell}$ are the same with Eq.(\ref{analytical2}) with same perihelion in Eq.(\ref{analytical-ph}) as shown in Model I.
  \end{enumerate}

\item Model IV: this corresponds to the cases with $\frac{1}{2}<k<1,M\leq0$, which have no horizon. One can take the right plots of Fig.\ref{geodesics1} for an example. We find that the meta-critical subcase, i.e. the one with $\Xi=\frac{1}{\ell}$, always has the positive effective potential. Hence this meta-critical subcase has no allowed motion, other than the unbounded spiral motions as other Models. When the constant of motions $\Xi$ changes, all the geodesic motions of a photon can be divided into two subcases:

  \begin{enumerate}
    \item when $0\leq\Xi\leq\frac{1}{\ell}$, the positive effective potential indicates that there is no allowed motion.
    \item when $\Xi>\frac{1}{\ell}$, the photon has the unbounded spiral motions. We calculate the numerical results for case with $\Xi=1.2$ in the right plot of Fig.\ref{geodesics1}: the perihelion is $r_{ph}\simeq1.7597$.
  \end{enumerate}

\item Model V: this corresponds to the cases with $k>1,M>0$. These corresponding metric have a single horizon. An example is presented in the right plots of Fig.\ref{geodesics2}. When the constant of motions $\Xi$ changes, the geodesic motions of a photon are all the unbounded spiral motions. For subcases with $0\leq\Xi<\frac{1}{\ell}$ and $\Xi\geq\frac{1}{\ell}$, there is still one difference: the latter one has no perihelion, thus the photon for the latter one will fall into the black holes. Follow the similar discussion with the radial geodesics of un-rotating photon in the previous subsection, one can find that the photon with $\Xi\geq\frac{1}{\ell}$ also arrive the black hole horizon in an infinite coordinate time.

\end{itemize}

\vspace{0.1in}

Considering all the above cases together, we present totally the geodesics of solutions with different mass $M$ in Tab.(\ref{tab1}) and Tab.(\ref{tab2}) corresponding to $k\in(\frac{1}{2},1)$ and $k\in(1,+\infty)$, respectively.   Comparing with the null
geodesics of the three dimensional regular charged black hole with $k=1$ \cite{Park:1999nc,Fernando:2003gg},
that of the EPM solutions with scalar hair have extra Model V which only contains the unbounded spiral motions.

\begin{table}[!htbp]
\centering
\begin{tabular}{|c|c|c|c|c|}
\hline

Solutions~&~  $M\leq0$ ~&~   $M\in(0,M_{ex})$ ~&~ $M=M_{ex}$ ~&~
$M>M_{ex}$  \\
\hline
Geodesics ~&~ Model IV ~&~  Model I ~&~  Model II ~&~  Model III  \\
\hline
\end{tabular}
\caption{The models of geodesic for solutions with $k\in(\frac{1}{2},1)$.}
\label{tab1}
\end{table}

\begin{table}[!htbp]
\centering
\begin{tabular}{|c|c|c|c|c|}
\hline

Solutions~&~  $M<M_{ex}$ ~&~ $M=M_{ex}$   ~&~ $M\in(M_{ex},0)$ ~&~
$M\geq0$   \\
\hline
Geodesics ~&~ Model I ~&~  Model II ~&~  Model III ~&~  Model V  \\
\hline
\end{tabular}
\caption{The models of geodesic for solutions with $k\in(1,+\infty)$.}
\label{tab2}
\end{table}

\section{Closing remarks}
\label{4s}
In this paper, we have presented an exact static solution to Einstein-Power-Maxwell theory in
$(2+1)$ dimensional AdS spacetimes, in which the scalar field couples to gravity in a non-minimal way. We focus on the simplified form of scalar field, then after considering the scalar potential, we have found that a stable system leads to a constraint on the power parameter $k$ of Maxwell field.
The solution contains a curvature singularity at the origin and is non-conformally flat.
The horizon structure are identified, then the existence of black hole solutions leads
to the physically acceptable mass bound $M\geq\,M_{ex}$, where $M_{ex}$ is the mass of the extreme black holes. Especially for the cases $k>1$, the lower bound $M_{ex}$ is negative, which shows that there exist scalar black holes with negative mass in three dimensions.

The geodesic motions for a
photon in this spacetime have been discussed in detail. They are divided into five models,
which are made up of the cases with the following geodesic motions: no-allowed motion, the circular motion,
the elliptic motion and the unbounded spiral motion. We also present some analytical results for the unbounded spiral motion. Differing from the null
geodesics of the three dimensional regular charged black hole with $k=1$ \cite{Park:1999nc,Fernando:2003gg},
that of the EPM black holes with scalar hair have extra Model V which only contains the unbounded spiral motions.
To consider the geodesics of this spacetime further, the time-like
one for a photon, even the one for charged particle are left as a future task. On the other hand, it would be also interesting to study the cause structure and thermodynamics of this black hole solution, especially for the one with negative mass.

\section*{Acknowledgements}

We would like to thank Liu Zhao for useful conversations. Wei Xu is supported by the
Research Innovation Fund of Huazhong University of Science and Technology (2014TS125).
This work was supported by the National Natural Science Foundation of China.


\begin{thebibliography}{100}

\bibitem{Bocharova}
 N.~M.~Bocharova,  K.~A.~Bronnikov, V.~N.~Melnikov,
``An exact solution of the system of Einstein equations and mass-free scalar field",
Vestn[J]. Mosk. univ. Fiz. astron, 1970, 6: 706.

\bibitem{Bekenstein:1974sf}
  J.~D.~Bekenstein,
  ``Exact solutions of Einstein conformal scalar equations,''
  Annals Phys.\  {\bf 82}, 535 (1974).

\bibitem{Bekenstein:1975ts}
  J.~D.~Bekenstein,
  ``Black Holes with Scalar Charge,''
  Annals Phys.\  {\bf 91}, 75 (1975).

\bibitem{Bronnikov:1978mx}
  K.~A.~Bronnikov and Y.~.N.~Kireev,
  ``Instability of Black Holes with Scalar Charge,''
  Phys.\ Lett.\ A {\bf 67}, 95 (1978).

\bibitem{Martinez:2004nb}
  C.~Martinez, R.~Troncoso and J.~Zanelli,
  ``Exact black hole solution with a minimally coupled scalar field,''
  Phys.\ Rev.\ D {\bf 70}, 084035 (2004)
  [\eprint{hep-th/0406111}]

\bibitem{Martinez:2006an}
  C.~Martinez and R.~Troncoso,
  ``Electrically charged black hole with scalar hair,''
  Phys.\ Rev.\ D {\bf 74}, 064007 (2006)  [\eprint{hep-th/0606130}].

\bibitem{Martinez:2005di}
  C.~Martinez, J.~P.~Staforelli and R.~Troncoso,
  ``Topological black holes dressed with a conformally coupled scalar field and electric charge,''
  Phys.\ Rev.\ D {\bf 74}, 044028 (2006)  [\eprint{hep-th/0512022}].

\bibitem{Nadalini:2007qi}
  M.~Nadalini, L.~Vanzo and S.~Zerbini,
  ``Thermodynamical properties of hairy black holes in n spacetimes dimensions,''
  Phys.\ Rev.\ D {\bf 77}, 024047 (2008)
  [\eprint{0710.2474}].

\bibitem{Kolyvaris:2009pc}
  T.~Kolyvaris, G.~Koutsoumbas, E.~Papantonopoulos and G.~Siopsis,
  ``A New Class of Exact Hairy Black Hole Solutions,''
  Gen.\ Rel.\ Grav.\  {\bf 43}, 163 (2011)  [\eprint{0911.1711}].

\bibitem{Gonzalez:2013aca}
  P.~A.~Gonz¨¢lez, E.~Papantonopoulos, J.~Saavedra and Y.~V¨¢squez,
  ``Four-Dimensional Asymptotically AdS Black Holes with Scalar Hair,''
  JHEP {\bf 1312}, 021 (2013)  [\eprint{1309.2161}].

\bibitem{Feng:2013tza}
  X.~-H.~Feng, H.~Lu and Q.~Wen,
  ``Scalar Hairy Black Holes in General Dimensions,''
  Phys.\ Rev.\ D {\bf 89}, 044014 (2014)  [\eprint{1312.5374}].

\bibitem{Acena:2012mr}
  A.~Acena, A.~Anabalon and D.~Astefanesei,
  ``Exact hairy black brane solutions in $AdS_{5}$ and holographic RG flows,''
  Phys.\ Rev.\ D {\bf 87}, no. 12, 124033 (2013)
  [\eprint{1211.6126}].

\bibitem{Acena:2013jya}
  A.~Ace?a, A.~Anabal¨®n, D.~Astefanesei and R.~Mann,
  ``Hairy planar black holes in higher dimensions,''
  JHEP {\bf 1401}, 153 (2014)
  [\eprint{1311.6065}].

\bibitem{Anabalon:2013sra}
  A.~Anabal¨®n and D.~Astefanesei,
  ``On attractor mechanism of $AdS_{4}$ black holes,''
  Phys.\ Lett.\ B {\bf 727}, 568 (2013)
  [\eprint{1309.5863}].

\bibitem{Anabalon:2012dw}
  A.~Anabalon,
  ``Exact Hairy Black Holes,''
  [\eprint{1211.2765}].
  
\bibitem{Herdeiro:2014goa}
  C.~A.~R.~Herdeiro and E.~Radu,
  ``Kerr black holes with scalar hair,''
  Phys.\ Rev.\ Lett.\  {\bf 112}, 221101 (2014)  [\eprint{1403.2757}].

\bibitem{Herdeiro:2014jaa}
  C.~Herdeiro and E.~Radu,
  ``Ergo-spheres, ergo-tori and ergo-Saturns for Kerr black holes with scalar hair,''
  [\eprint{1406.1225}].

\bibitem{Gaete:2013ixa}
  M.~Bravo Gaete and M.~Hassaine,
  ``Topological black holes for Einstein-Gauss-Bonnet gravity with a nonminimal scalar field,''
  Phys.\ Rev.\ D {\bf 88}, 104011 (2013)  [\eprint{1308.3076}].

\bibitem{Gaete:2013oda}
  M.~Bravo Gaete and M.~Hassaine,
  ``Planar AdS black holes in Lovelock gravity with a nonminimal scalar field,''
  JHEP {\bf 1311}, 177 (2013)  [\eprint{1309.3338}].

\bibitem{Correa:2013bza}
  F.~Correa and M.~Hassaine,
  ``Thermodynamics of Lovelock black holes with a nonminimal scalar field,''
  JHEP {\bf 1402}, 014 (2014)  [\eprint{1312.4516}].

\bibitem{Giribet:2014bva}
  G.~Giribet, M.~Leoni, J.~Oliva and S.~Ray,
  ``Hairy black holes sourced by a conformally coupled scalar field in D dimensions,''
  Phys.\ Rev.\ D {\bf 89}, 085040 (2014)  [\eprint{1401.4987}].

\bibitem{Banados:1992wn}
  M.~Banados, C.~Teitelboim and J.~Zanelli,
  ``The Black hole in three-dimensional space-time,''
  Phys.\ Rev.\ Lett.\  {\bf 69}, 1849 (1992)
  [\eprint{hep-th/9204099}].

\bibitem{Cataldo:1999wr}
  M.~Cataldo and A.~Garcia,
  ``Three dimensional black hole coupled to the Born-Infeld electrodynamics,''
  Phys.\ Lett.\ B {\bf 456}, 28 (1999)
  [\eprint{hep-th/9903257}].

\bibitem{Mazharimousavi:2011nd}
  S.~H.~Mazharimousavi, O.~Gurtug, M.~Halilsoy and O.~Unver,
  ``2+1 dimensional magnetically charged solutions in Einstein - Power - Maxwell theory,''
  Phys.\ Rev.\ D {\bf 84}, 124021 (2011)
  [\eprint{1103.5646}].

\bibitem{Gurtug:2010dr}
  O.~Gurtug, S.~H.~Mazharimousavi and M.~Halilsoy,
  ``2+1-dimensional electrically charged black holes in Einstein - power Maxwell Theory,''
  Phys.\ Rev.\ D {\bf 85}, 104004 (2012)
  [\eprint{1010.2340}].

\bibitem{Chan:1994qa}
  K.~C.~K.~Chan and R.~B.~Mann,
  Phys.\ Rev.\ D {\bf 50}, 6385 (1994)
  [Erratum-ibid.\ D {\bf 52}, 2600 (1995)]
  [\eprint{gr-qc/9404040}].

\bibitem{Martinez:1999qi}
  C.~Martinez, C.~Teitelboim and J.~Zanelli,
  Phys.\ Rev.\ D {\bf 61}, 104013 (2000)
  [\eprint{hep-th/9912259}].

\bibitem{Astorino:2011mw}
  M.~Astorino,
  ``Accelerating black hole in 2+1 dimensions and 3+1 black (st)ring,''
  JHEP {\bf 1101}, 114 (2011)
  [\eprint{1101.2616}].  

\bibitem{Xu:2011vp}
  W.~Xu, K.~Meng and L.~Zhao,
  ``Accelerating BTZ spacetime,''
  Class.\ Quant.\ Grav.\  {\bf 29}, 155005 (2012)
  [\eprint{1111.0730}].  

\bibitem{Garcia:2002rn}
  A.~A.~Garcia and C.~Campuzano,
  ``All static circularly symmetric perfect fluid solutions of (2+1) gravity,''
  Phys.\ Rev.\ D {\bf 67}, 064014 (2003)
  [\eprint{gr-qc/0211014}].  

\bibitem{Wu:2013wca}
  B.~Wu and W.~Xu,
  ``New class of rotating perfect fluid black holes in three dimensional gravity,''
  Eur.\ Phys.\ J.\ C {\bf74}, 3007 (2014)
  [\eprint{1312.6741}].  

\bibitem{Hassaine:2007py}
  M.~Hassaine and C.~Martinez,
  ``Higher-dimensional black holes with a conformally invariant Maxwell source,''
  Phys.\ Rev.\ D {\bf 75}, 027502 (2007)  [\eprint{hep-th/0701058}].

\bibitem{Hassaine:2008pw}
  M.~Hassaine and C.~Martinez,
  ``Higher-dimensional charged black holes solutions with a nonlinear electrodynamics source,''
  Class.\ Quant.\ Grav.\  {\bf 25}, 195023 (2008)  [\eprint{0803.2946}].

\bibitem{Maeda:2008ha}
  H.~Maeda, M.~Hassaine and C.~Martinez,
  ``Lovelock black holes with a nonlinear Maxwell field,''
  Phys.\ Rev.\ D {\bf 79}, 044012 (2009)  [\eprint{0812.2038}].

\bibitem{DiazAlonso:2012mb}
  J.~Diaz-Alonso and D.~Rubiera-Garcia,
  ``Thermodynamic analysis of black hole solutions in gravitating nonlinear electrodynamics,''
    Gen.\  Rel.\  Grav.\  {\bf 45}, 1901 (2013)  [\eprint{1204.2506}].

\bibitem{Gonzalez:2009nn}
  H.~A.~Gonzalez, M.~Hassaine and C.~Martinez,
  ``Thermodynamics of charged black holes with a nonlinear electrodynamics source,''
    Phys.\ Rev.\ D {\bf 80}, 104008 (2009)  [\eprint{0909.1365}].

\bibitem{Bazrafshan:2012rn}
  A.~Bazrafshan, M.~H.~Dehghani and M.~Ghanaatian,
  ``Surface Terms of Quartic Quasitopological Gravity and Thermodynamics of Nonlinear Charged Rotating Black Branes,''
    Phys.\ Rev.\ D {\bf 86}, 104043 (2012)  [\eprint{1209.0246}].

\bibitem{Arciniega:2014iya}
  G.~Arciniega and A.~S¨¢nchez,
  ``Geometric description of the thermodynamics of a black hole with power Maxwell invariant source,''
    [\eprint{1404.6319}].

\bibitem{Rasheed:1997ns}
  D.~A.~Rasheed,
  ``Nonlinear electrodynamics: Zeroth and first laws of black hole mechanics,''
    [\eprint{hep-th/9702087}].

\bibitem{Hendi:2012um}
  S.~H.~Hendi and M.~H.~Vahidinia,
  ``Extended phase space thermodynamics and P-V criticality of black holes with a nonlinear source,''
    Phys.\ Rev.\ D {\bf 88}, no. 8, 084045 (2013) [\eprint{1212.6128}].

\bibitem{Mo:2014qsa}
  J.~X.~Mo and W.~B.~Liu,
  ``$P-V$ criticality of topological black holes in Lovelock-Born-Infeld gravity,''
   Eur.\ Phys.\ J.\ C {\bf 74}, 2836 (2014)  [\eprint{1401.0785}].

\bibitem{Banados:2005hm}
  M.~Banados and S.~Theisen,
  ``Scale invariant hairy black holes,''
  Phys.\ Rev.\ D {\bf 72}, 064019 (2005)
  [\eprint{hep-th/0506025}].

\bibitem{Henneaux:2002wm}
  M.~Henneaux, C.~Martinez, R.~Troncoso and J.~Zanelli,
  ``Black holes and asymptotics of 2+1 gravity coupled to a scalar field,''
  Phys.\ Rev.\ D {\bf 65}, 104007 (2002)
  [\eprint{hep-th/0201170}].

\bibitem{Schmidt:2012pp}
  H.~J.~Schmidt and D.~Singleton,
  ``Exact radial solution in 2+1 gravity with a real scalar field,''
  Phys.\ Lett.\ B {\bf 721}, 294 (2013)
  [\eprint{1212.1285}].

\bibitem{Hortacsu:2003we}
  M.~Hortacsu, H.~T.~Ozcelik and B.~Yapiskan,
  ``Properties of solutions in (2+1)-dimensions,''
  Gen.\ Rel.\ Grav.\  {\bf 35}, 1209 (2003)
  [\eprint{gr-qc/0302005}].

\bibitem{Martinez:1996gn}
  C.~Martinez and J.~Zanelli,
  ``Conformally dressed black hole in (2+1)-dimensions,''
  Phys.\ Rev.\ D {\bf 54}, 3830 (1996)  [\eprint{gr-qc/9604021}].

\bibitem{Xu:2013nia}
  W.~Xu and L.~Zhao,
  ``Charged black hole with a scalar hair in $(2+1)$ dimensions,''
  Phys.\ Rev.\ D {\bf 87}, 124008 (2013)
  [\eprint{1305.5446}].

\bibitem{Zhao:2013isa}
  L.~Zhao, W.~Xu and B.~Zhu,
  ``Novel rotating hairy black hole in $(2+1)$-dimensions,''
  Commun.\ Theor.\ Phys.\  {\bf 61}, 475 (2014)
  [\eprint{1305.6001}].

\bibitem{Degura:1998hw}
  Y.~Degura, K.~Sakamoto and K.~Shiraishi,
  ``Black holes with scalar hair in (2+1)-dimensions,''
  Grav.\ Cosmol.\  {\bf 7}, 153 (2001)
  [\eprint{gr-qc/9805011}].

\bibitem{Aparicio:2012yq}
  J.~Aparicio, D.~Grumiller, E.~Lopez, I.~Papadimitriou and S.~Stricker,
  ``Bootstrapping gravity solutions,''
  JHEP {\bf 1305}, 128 (2013)
  [\eprint{1212.3609}].

\bibitem{Zou:2014gla}
  D.~C.~Zou, Y.~Liu, B.~Wang and W.~Xu,
  ``Rotating black holes with scalar hair in three dimensions,''  
  [\eprint{1408.2419}].  

\bibitem{Sadeghi:2013gmf}
  J.~Sadeghi, B.~Pourhassan and H.~Farahani,
  ``Rotating charged hairy black hole in (2+1) dimensions and particle acceleration,''
  [\eprint{1310.7142}].

\bibitem{Mazharimousavi:2014vza}
  S.~H.~Mazharimousavi and M.~Halilsoy,
  ``Einstein-Born-Infeld black holes with a scalar hair in three-dimensions,''
  [\eprint{1405.2956}].  

\bibitem{Xu:2014uha}
  W.~Xu, L.~Zhao and D.~-C.~Zou,
  ``Three dimensional rotating hairy black holes, asymptotics and thermodynamics,''
  [\eprint{1406.7153}].

\bibitem{Correa:2014ika}
  F.~Correa, M.~Hassaine and J.~Oliva,
  ``Black holes in New Massive Gravity dressed by a (non)minimally coupled scalar field,''
  Phys.\ Rev.\ D {\bf 89}, 124005 (2014)  [\eprint{1403.6479}].

\bibitem{Jing:2011vz}
  J.~Jing, Q.~Pan and S.~Chen,
  ``Holographic Superconductors with Power-Maxwell field,''
  JHEP {\bf 1111}, 045 (2011)  [\eprint{1106.5181}].  

\bibitem{Jing:2012dj}
  J.~Jing, Q.~Pan and S.~Chen,
  ``Holographic Superconductor/Insulator Transition with logarithmic electromagnetic field in Gauss-Bonnet gravity,''
    Phys.\ Lett.\ B {\bf 716}, 385 (2012)  [\eprint{1209.0893}].

\bibitem{Banerjee:2012vk}
  R.~Banerjee, S.~Gangopadhyay, D.~Roychowdhury and A.~Lala,
  ``Holographic s-wave condensate with non-linear electrodynamics: A nontrivial boundary value problem,''
  Phys.\ Rev.\ D {\bf 87}, 104001 (2013)  [\eprint{1208.5902}].

\bibitem{Roychowdhury:2012vj}
  D.~Roychowdhury,
  ``AdS/CFT superconductors with Power Maxwell electrodynamics: reminiscent of the Meissner effect,''
    Phys.\ Lett.\ B {\bf 718}, 1089 (2013)  [\eprint{1211.1612}].

\bibitem{Dey:2013qoa}
  S.~Dey and A.~Lala,
  ``Holographic $s$-wave condensation and Meissner-like effect in Gauss-Bonnet gravity with various non-linear corrections,''
    [\eprint{1306.5137}].  
 
\bibitem{Chan:2008gv}
  C.~-k.~Chan,
  ``Oscillations of the Inner Regions of Viscous Accretion Disks,''
  Astrophys.\ J.\  {\bf 704}, 68 (2009)  [\eprint{0812.2031}].

\bibitem{Cardenas}
  M.~Cardenas, O.~Fuentealba and C.~Martinez,
  ``Three-dimensional black holes with conformally coupled scalar and gauge fields,''
  [\eprint{1408.1401}].  

\bibitem{Ida:2000jh}
  D.~Ida,
  ``No black hole theorem in three-dimensional gravity,''
  Phys.\ Rev.\ Lett.\  {\bf 85}, 3758 (2000)  [\eprint{gr-qc/0005129}].

\bibitem{Brown:1992br}
  J.~D.~Brown and J.~W.~York, Jr.,
  ``Quasilocal energy and conserved charges derived from the gravitational action,''
  Phys.\ Rev.\ D {\bf 47}, 1407 (1993)
  [\eprint{gr-qc/9209012}].

\bibitem{Brown:1994gs}
  J.~D.~Brown, J.~Creighton and R.~B.~Mann,
  ``Temperature, energy and heat capacity of asymptotically anti-de Sitter black holes,''
  Phys.\ Rev.\ D {\bf 50}, 6394 (1994)
  [\eprint{gr-qc/9405007}].

\bibitem{Creighton:1995au}
  J.~D.~E.~Creighton and R.~B.~Mann,
  ``Quasilocal thermodynamics of dilaton gravity coupled to gauge fields,''
  Phys.\ Rev.\ D {\bf 52}, 4569 (1995)
  [\eprint{gr-qc/9505007}].

\bibitem{Vanzo:1997gw}
  L.~Vanzo,
  ``Black holes with unusual topology,''
  Phys.\ Rev.\ D {\bf 56}, 6475 (1997)
  [\eprint{gr-qc/9705004}].

\bibitem{Mann:1997jb}
  R.~B.~Mann,
  ``Black holes of negative mass,''
  Class.\ Quant.\ Grav.\  {\bf 14}, 2927 (1997)
  [\eprint{gr-qc/9705007}].

\bibitem{Brill:1997mf}
  D.~R.~Brill, J.~Louko and P.~Peldan,
  ``Thermodynamics of (3+1)-dimensional black holes with toroidal or higher genus horizons,''
  Phys.\ Rev.\ D {\bf 56}, 3600 (1997)
  [\eprint{gr-qc/9705012}].

\bibitem{Belletete:2013nqa}
  J.~Bellet¨ºte and M.~B.~Paranjape,
  ``On negative mass,''
  Int.\ J.\ Mod.\ Phys.\ D {\bf 22}, 1341017 (2013)
  [\eprint{1304.1566}].

\bibitem{Mbarek:2014ppa}
  S.~Mbarek and M.~B.~Paranjape,
  ``Negative mass bubbles in de Sitter space-time,''
  [\eprint{1407.1457}].

\bibitem{Park:1999nc}
  D.~H.~Park and S.~H.~Yang,
  ``Geodesic motions in (2+1)-dimensional charged black holes,''
  Gen.\ Rel.\ Grav.\  {\bf 31}, 1343 (1999)
  [\eprint{gr-qc/9901027}].

\bibitem{Fernando:2003gg}
  S.~Fernando, D.~Krug and C.~Curry,
  ``Geodesic structure of static charged black hole solutions in 2+1 dimensions,''
  Gen.\ Rel.\ Grav.\  {\bf 35}, 1243 (2003).

\end{thebibliography}

\providecommand{\href}[2]{#2}\begingroup
\footnotesize\itemsep=0pt
\providecommand{\eprint}[2][]{\href{http://arxiv.org/abs/#2}{arXiv:#2}}

\endgroup

\end{document}